\documentclass[usenatbib]{mn2e}
\usepackage{graphicx,amsmath,amssymb}

\begin{document}
\topmargin-1cm

\newcommand\approxgt{\mbox{$^{>}\hspace{-0.24cm}_{\sim}$}}
\newcommand\approxlt{\mbox{$^{<}\hspace{-0.24cm}_{\sim}$}}
\newcommand{\be}{\begin{equation}}
\newcommand{\ee}{\end{equation}}
\newcommand{\bea}{\begin{eqnarray}}
\newcommand{\eea}{\end{eqnarray}}
\newcommand{\lexp}{\mathop{\langle}}
\newcommand{\rexp}{\mathop{\rangle}}
\newcommand{\rexpc}{\mathop{\rangle_c}}
\newcommand{\nbar}{\bar{n}}
\newcommand{\zmax}{Z_{\rm max}}
\newcommand{\amass}{10^{13} \,M_\odot}
\newcommand{\hmass}{10^{14} \,M_\odot}
\newcommand{\mmin}{M_{\rm{min}}}

\title{Cluster Galaxy Dynamics and the Effects of Large Scale Environment}

\author[White, Cohn \& Smit]{Martin White${}^{1,2}$, J.D.~Cohn${}^{3}$
and Renske Smit${}^{2,4}$\\
${}^1$ Department of Physics, University of California, Berkeley, CA 94720\\
${}^2$ Department of Astronomy, University of California, Berkeley, CA 94720\\
${}^3$ Space Science Laboratory, University of California, Berkeley, CA 94720\\
${}^4$ Sterrewacht Leiden, Postbus 9513, 2300 RA Leiden, the Netherlands}

\date{\today}

\pagerange{\pageref{firstpage}--\pageref{lastpage}} \pubyear{2010}

\maketitle

\label{firstpage}

\begin{abstract}
Advances in observational capabilities have ushered in a new era of
multi-wavelength, multi-physics probes of galaxy clusters and ambitious
surveys are compiling large samples of cluster candidates selected in
different ways.
We use a high-resolution N-body simulation to study how the
influence of large-scale structure in and around clusters causes correlated
signals in different physical probes and discuss some implications this
has for multi-physics probes of clusters (e.g.~richness, lensing,
Compton distortion and velocity dispersion).  

We pay particular attention to velocity dispersions, matching
galaxies to subhalos which are explicitly tracked in the simulation.
We find that not only do halos persist as subhalos when they fall into a
larger host, groups of subhalos retain their identity for long periods within
larger host halos.
The highly anisotropic nature of infall into massive clusters, and their
triaxiality, translates into an anisotropic velocity ellipsoid:
line-of-sight galaxy velocity dispersions for any individual halo show large
variance depending on viewing angle.
The orientation of the velocity ellipsoid is correlated with the large-scale
structure, and thus velocity outliers correlate with outliers caused by
projection in other probes.
We quantify this orientation uncertainty and give illustrative examples.
Such a large variance suggests that velocity dispersion estimators will
work better in an ensemble sense than for any individual cluster, which
may inform strategies for obtaining redshifts of cluster members.
We similarly find that the ability of substructure indicators to find
kinematic substructures is highly viewing angle dependent.
While groups of subhalos which merge with a larger host halo can retain
their identity for many Gyr, they are only sporadically picked up by
substructure indicators.

We discuss the effects of correlated scatter on scaling relations estimated
through stacking, both analytically and in the simulations, showing that
the strong correlation of measures with mass and the large scatter in mass
at fixed observable mitigate line-of-sight projections.
\end{abstract}

\section{Introduction}

Galaxy clusters form the high-mass tail of hierarchical structure formation
and are of interest for constraining cosmological parameters, understanding
large scale structure, as extreme environments for galaxy formation and as
objects hosting unique astrophysical phenomena.
While first discovered as concentrations of galaxies \citep{Abe58,Zwi66},
they are now also routinely found as luminous, extended X-ray sources
\citep{Sch78,McH78,Boh00,Boh01},
as peaks in the shear field \citep{Wit06} and as ``holes'' in the microwave sky
\citep{SPT}.
To mitigate the systematic errors associated with each individual method and
to provide a more complete understanding of clusters, multi-wavelength studies
have become increasingly common.  Each waveband adds knowledge about clusters.
However, we might expect there to be significant correlations between effects
in different methods both because the intrinsic properties they measure depend
on e.g.~cluster size but also because they are similarly affected by the
complex environment surrounding clusters.  

\begin{figure}
\begin{center}
\resizebox{4in}{!}{\includegraphics{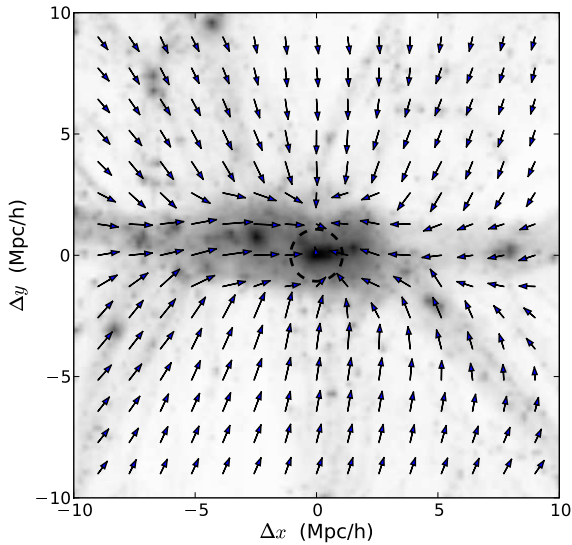}}
\end{center}
\caption{The velocity field traces the filamentary large-scale structure.
The grey scale shows the logarithm of the projected dark matter density in a
$5\,h^{-1}$Mpc thick slice around a cluster of mass
$5\times 10^{14}\,h^{-1}M_\odot$ at $z=0$.
The slice is oriented to contain the halo particle velocity eigenvectors with
the largest ($x$) and smallest ($y$) eigenvalues.
The dashed circle shows the region within $r_{200c}$.
Note that within the cluster the velocities trace the elongation of the
matter, as required.  In addition the velocities at larger distances trace
the filamentary structure in the larger scale environment.  Thus we expect
that velocity anisotropy will be correlated with density anisotropy.}
\label{fig:dotplot}
\end{figure}

Roughly speaking, both the hot gas and galaxies in clusters trace the dark
matter which dominates the potential.
We can approximate the clusters as self-similar and isothermal, with a
temperature $T\propto M^{2/3}$, a velocity dispersion $\sigma^2\propto T$, and
richness $N\propto M$ for sufficiently massive halos \citep{Kai86}.
Consider a small region near the cluster: lensing measures the sum of all
of the mass, richness all of the mass in halos above some threshold and 
Compton (or SZ) distortion \citep[][SZ]{SunZel72} all of the mass in halos
weighted by $M^{2/3}$.
The signal in each probe depends on the mass within the virialized region of
the cluster, the mass near the cluster but outside the virial region and
(uncorrelated) mass at larger distance along the line-of-sight.
Lensing and Compton distortion measures provide little line-of-sight resolution.
The degree of projection involved in a richness estimator depends on how well
the galaxy distances are known (e.g.~using photometric or spectroscopic
redshifts).  

Line-of-sight galaxy velocities in principle provide a measure of the potential well
depth or mass and offer the possibility of breaking line-of-sight projection.
However, the velocity field traces the density field, and can be correlated
with line-of-sight projection due to the filamentary nature of mass accretion
onto massive halos (see Fig.~\ref{fig:dotplot} which gives an example of this
effect in our simulations).
The velocities of cluster galaxies can retain this large scale anisotropy
\citep[see also][for studies of dark matter velocities]{Tor97,KasEvr05}.
Thus it is easy to imagine that line-of-sight velocity dispersion could be
correlated with filamentary material which can bias individual cluster
measurements in e.g.~richness, lensing or Compton distortion. 

We would like to investigate
how the complex structure of the cosmic web of material near clusters leads
to correlations in individual cluster observables, and the implications that
this has for these four probes of clusters.
This shared dependence, not only on cluster properties but also on cluster
environment, can introduce additional subtleties when methods are combined.
For example, an often used approach is to ``stack'' clusters on the basis of
one observed property $X$ (e.g.~richness), and then look for correlations
between two other properties, $Y$ and $Z$.
Clearly, it is very important to understand the joint distribution $P(X,Y,Z)$
and the degree of correlation between scatter in $X$, $Y$ or $Z$.

In this paper we use N-body numerical simulations with subhalos (which we
identify with galaxies) to study properties of cluster galaxy kinematics and
the relation of the scatters in velocity dispersion, Compton distortion,
lensing, and optical richness\footnote{We do not address X-ray emission
in this paper.} generated by nearby large-scale structure.
Details of our numerical simulations and methods for finding subhalos
are given in \S\ref{sec:sims}.  We describe how the mock richness, lensing and
Compton distortion observations are constructed in \S\ref{sec:mock}.
Readers interested in the results may skip to \S\ref{sec:intrinsic} where
we discuss the intrinsic properties of our massive halos and their subhalo (galaxy)
populations and \S\ref{sec:interlopers} where we discuss measurements of galaxy kinematics
in the presence of interlopers and \S\ref{sec:observed} where we discuss
the correlations between different observables.

The effects of the cosmic web, and in particular projection effects, have been
long-time concerns for optical cluster finding
\citep[e.g.][]{Abe58,Dal92,Lum92,vHaarlem97,Whi99}, 
measuring Compton distortion
\citep[e.g.][]{WhiHerSpr02,HolMcCBab07,Hal07,ShaHolBod08},
or interpreting weak lensing maps
\citep[e.g.][]{RebBar99,MetWhiLok01,Hoe01,PutWhi05,Men10}.   
Correlations between scatters induced by common projection effects were noted
in \citet{CohWhi09}. \citet{Cen97}
did an early simulation study of projection on several of the indicators we consider here
including richness, velocity dispersions and lensing, and measured substructure
using dark matter particles.
For cluster kinematics in particular, the velocity dispersion properties of
dark matter particles and their relation to the cosmic web were studied in
\citet{Tor97,KasEvr05} and \citet{Biv06} noted that filamentary inflow was
expected to affect measured velocity dispersions.
Our simulations have enough dynamic range that we can simulate a representative
cosmological volume, including the neighboring large-scale structure and
cosmic web, while simultaneously resolving and tracking the subhalos which we
believe are galactic hosts.  This preserves any correlations between subhalo
properties and halo orientation or cosmic web, and coherence between subhalo
populations which fell in as part of a group.
We emphasize the effect that anisotropy in galaxy kinematics has on
line-of-sight velocity dispersion or virial mass estimators of cluster mass
and discuss how this scatter compares to (and correlates with) other measures
of cluster mass which are sensitive to the cluster's environment.
This forms a partial extension of the work of \citet{Sta10}, who discussed
the correlations in scatters of a large number of different intrinsic
(rather than projected) cluster quantities, including X-ray.

\section{Simulations} \label{sec:sims}

In order to investigate the above questions with `realistic' conditions we
need mock galaxy, gas and lensing catalogs in which clusters of galaxies are placed
in their correct cosmological context, with an appropriate prescription for
identifying galaxies and for which the intrinsic cluster properties are known.
We make use of several dark-matter-only N-body simulations.
Such simulations follow the evolution of large dark matter halos, which
we observe as galaxy clusters, correctly accounting for their place in
the filamentary large-scale structure and their complex formation histories.

\subsection{N-body simulation}

We make use of several simulations in this paper.  The main one is of the
$\Lambda$CDM family with $\Omega_m=0.274$, $\Omega_\Lambda=0.726$,
$h=0.7$, $n=0.95$ and $\sigma_8=0.8$, in agreement with a wide array of
observations.
Briefly, we used the TreePM code described in \citet{TreePM} to evolve $2048^3$
equal mass particles in a periodic cube of side length $250\,h^{-1}$Mpc.
This results in particle masses of $1.4\times 10^8\,h^{-1}M_\odot$ and a
Plummer equivalent smoothing of $2.5\,h^{-1}$kpc.
The initial conditions were generated by displacing particles from a regular
grid using second order Lagrangian perturbation theory at $z=150$ where the
rms displacement is $38$ per cent of the mean inter-particle spacing.
The phase space data were dumped at 45 times, equally spaced in $\ln(a)$ from
$z=10$ to $0$.
This TreePM code has been compared to a number of other codes and shown to
perform well for such simulations \citep{Hei08}.
Though we shall not highlight them individually, in addition to this
simulation we have made use of the simulation described in \citet{WetWhi10},
which used a different subhalo finder, and four other simulations of smaller
volumes focused on massive halos where we have mass resolution $2-5$ times
higher than in the fiducial run and comparably higher force resolution.
This allows us to check the dependence on subhalo finding and tracking
scheme, mass and force resolution and on limiting mass.

For each output we found dark matter halos using the Friends of Friends (FoF)
algorithm \citep{DEFW} with a linking length of $0.168$ times the mean
interparticle spacing.  This partitions the particles into equivalence
classes roughly bounded by isodensity contours of $100\times$ the mean
density.
We keep all halos above $50$ particles, and generate merger trees for all
of the halos in the simulation so as to identify the times of last major
mergers or other interesting events in the history.
The center of the halo is taken to be the position of the most bound particle,
including all of the mass in the Friends of Friends halo in the computation of
the potential.

Given halo centers, 
we also compute the spherically averaged mass profile taking into account
all of the mass in the simulation.
We follow standard convention and define the virial radius as that radius
within which the mean density is $200$ times the critical density at the
epoch of observation, writing this $r_{200c}$.
The three dimensional velocity dispersion of the dark matter within
$r_{200c}$ is tightly correlated with the mass interior to the same
radius as expected from the virial relation
\citep[e.g.][for a recent study]{Evr08}.
The mass function of halos is approximately universal if a density
contrast tied to the mean density and encompassing the zero-velocity
surface is used \citep{Jen01,Whi01,Rob09,Bha10}.
When appropriate we also use the radius within which the mean density is
$180$ times the background density, $r_{180b}$, for convenience.
Unless stated below, the mass quoted will be $M_{180b}$.

\subsection{Subhalos/Galaxies} \label{sec:galaxies}

In hierarchical structure formation models, such as CDM, the virialized
regions of large dark matter halos contain subhalos
--- self-gravitating, bound clumps of dark matter ---
which contain $\mathcal{O}(10)$ per cent of the total halo mass.
Luminous galaxies form via the cooling and condensation of baryons in the
very centers of halos and subhalos so these subhalos identify the sites of
galaxy formation in the simulation.

We identify ``subhalos'' within our Friends of Friends halos as overdensities
in phase space (see Appendix \ref{app:fof6d}).
For newly formed halos the ``central'' subhalo is defined as the most
massive subhalo within the host.  For other halos it is defined as the
descendant of the central subhalo within the most massive progenitor of
the host halo.  The subhalo position is that of its most bound particle.
Subhalo merger trees are computed as described in \citet{WetCohWhi09} and
\citet{WetWhi10}.
Briefly, subhalo histories are tracked across 4 consecutive output
times to ensure subhalos are not ``lost'' during close passes through the
dense central regions of a halo.  Parent-child relationships are determined
using the 20 most bound particles (which we have found to be very stable).
For each subhalo we define $M_{\rm inf}$ as the host halo mass it had just
prior to becoming a satellite, i.e.~the largest host halo mass for which
it was the central subhalo.
We shall use $M_{\rm inf}$ as a proxy for stellar mass or luminosity and
keep all subhalos whose infall mass is larger than
$2\times 10^{11}\,h^{-1}M_\odot$ ($>10^3$ particles).
Resolution tests indicate the catalogs are largely complete to this mass
limit (see also \citealt{MBK09}).

As discussed extensively in \citet{WetWhi10} there are slightly more satellites
per host halo, and correspondingly more small-scale clustering power, than
observations demand.  If we remove the excess based on the ratio of
instantaneous subhalo mass to infall mass \citep{WetWhi10} and match subhalos
to galaxies based on abundance then our halo catalog is in good agreement
with many observations including the global and cluster luminosity
functions, the satellite statistics and the luminosity dependent clustering
of galaxies.

We shall focus primarily on $z\simeq 0.1$, discussing what changes as we go
to higher redshift in \S\ref{sec:highz}.
At $z=0.1$, the number density of subhalos above our mass threshold is
$0.02 h^{-3}{\rm Mpc}^3$.
Observationally the same number densities are achieved  by going down to
$0.2\,L_\star$ or $M_r=-18.5$ in the r-band \citep{Bla03}
or about $M_B=-18.6$ \citep{Fab07}
or a stellar mass of about $3\times 10^9\,h^{-1}M_\odot$ \citep{Mos10}.

As our simulation explicitly tracks the evolution of subhalos, including
their complex dynamics and mass loss, we are in a position to ask
sophisticated questions about the spatial and kinematic distribution of
``galaxies'' in clusters.  
As shown below, the subhalo spatial distribution and its environment
dependence is in good agreement with corresponding observations of galaxies.
By using subhalos, rather than just randomly drawing particles from within
the halo, we ensure that we keep any correlations between subhalo positions
and the halo orientation or its large-scale environment
\citep[see e.g.][for recent reviews]{Fal09,Siv10}
and between positions and dynamics of subhalos that fell into the host as
part of a larger group.

\begin{figure}
\begin{center}
\resizebox{3in}{!}{\includegraphics{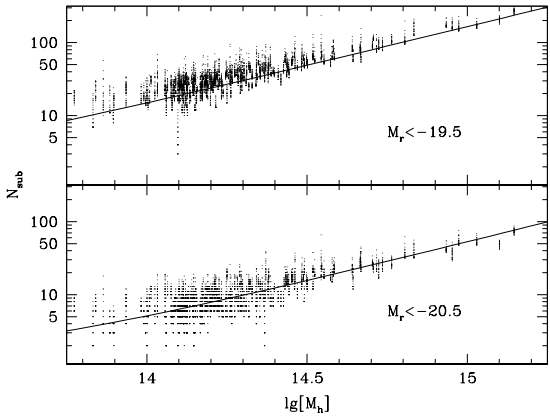}}
\end{center}
\caption{The halo occupation distribution for our clusters at $z\sim 0.1$
(points), compared to the SDSS group catalogs of
\protect\citet[][solid lines]{Yan08}
for two different luminosity thresholds.  Subhalo masses are converted to
$r$-band luminosity by abundance matching, assuming a 1-1 relation
(i.e.~no scatter).  The panels correspond to lg$M>11.7$ and 12.2
respectively, with masses measured in $h^{-1}M_\odot$.  For each cluster
the halo mass is defined as that interior to $r_{180b}$, within which the
mean matter density is 180 times the background density.  The richness is
defined as all galaxies above a phase-space density threshold, as in the
observations.  We plot 30 lines-of-sight per halo.}
\label{fig:yang}
\end{figure}

\begin{figure}
\begin{center}
\resizebox{3in}{!}{\includegraphics{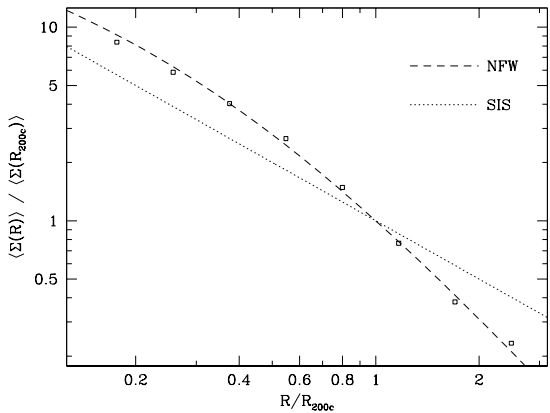}}
\end{center}
\caption{The (projected) subhalo profile for galaxies brighter than
$0.4\,L_\star$ (i.e.~$M_\star+1$), normalized at the virial radius.
The curves show a simple singular isothermal sphere ($\Sigma\propto R^{-1}$)
and the profile \protect\citet{LinMohSta04} found which fits the counts of
$K$-band selected galaxies at $z\simeq 0.1$.}
\label{fig:projdist}
\end{figure}

\begin{figure}
\begin{center}
\resizebox{3in}{!}{\includegraphics{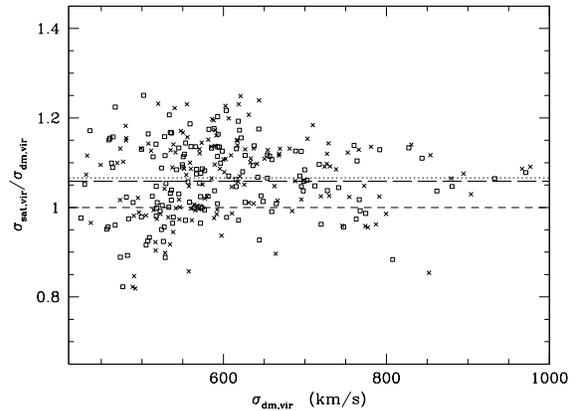}}
\end{center}
\caption{The isotropically averaged 1D velocity dispersion of the dark matter
vs.~the satellite subhalos for halos with more than 50 members at $z\simeq 0.1$.
The points are for all particles and satellites within spheres of radius
$r_{200c}$ (crosses) and $r_{100c}$ (squares) about the most bound particle
in the halo.
The short-dashed line represents equality and the dotted ($r_{200c}$) and
long-dashed ($r_{100c})$ lines are an unweighted fit to the points.}
\label{fig:sig3D}
\end{figure}

In Figure \ref{fig:yang} we demonstrate that the halo occupation distribution
of galaxies in the simulation is in good agreement with the measurements from
the group catalog in \citet{Yan08}.
Our satellite spatial distribution is slightly shallower than the dark
matter in the central regions.
The (projected) profile matches well the NFW profile found to fit
the counts of $K$-band selected galaxies by \citet{LinMohSta04} down to
$r\simeq 0.1\, r_{200c}$, see Figure \ref{fig:projdist}.

We find evidence for mild positive velocity bias within the virial sphere
(Fig.~\ref{fig:sig3D}), in agreement with most previous work
\citep{Gao04,Got05,FalDie06,LauNagKra10,Fal10}.
The velocities of satellites appears to be determined almost entirely by
the hosts' potential \citep{Wet10}, although we caution that the degree
of velocity bias does depend on the manner in which subhalos are selected
and retained, with more massive satellites showing reduced dispersion and
satellites accreted more recently generally having increased
dispersion\footnote{\citet{Biv06} also showed that the bias differed for
`early-type' and `late-type' subhalos.}.
This may become important as we move to higher redshift where the mean mass
of the host halos should decrease while the mean mass of the halos for which
one could obtain accurate redshifts should increase, leading to a smaller mass
ratio and larger effects of dynamical friction.
This is partially canceled by the ``more recent'' infall time distribution at
higher redshift.
In summary, the exact amount of velocity bias will depend on how the subhalo
samples are selected \citep{Gao04,Got05,FalDie06,LauNagKra10,Fal10},
and can evolve with redshift -- since it is not entirely clear how to match
an observed galaxy sample to a particular subhalo sample it seems prudent
to assign a $\mathcal{O}(10)$ per cent theoretical uncertainty in the absolute
value of the velocity bias with a roughly comparable scatter from halo to halo.

\subsection{Missing physics}

Our simulations do not attempt to model the baryonic
component, and thus can only be an approximation to the full story.
Fortunately for the most massive objects in the universe, the majority of the baryonic material is in hot gas, rather than cold gas or stars.
The cooling of gas in massive clusters does not dramatically alter the halo
profile, except in the very inner regions \citep[e.g.][]{Kaz04}
Outside of these regions the spatial distribution of the hot gas largely
follows the gravitational potential and we shall make this assumption where
necessary.
The hot intra-cluster medium in massive halos is expected to alter the
orbits of satellites only mildly \citep{Sim09,LauNagKra10,JiaJinLin10},
since it is a minority mass component and they are traveling at close to
the sound speed \cite[e.g.][]{ConOst08}.
The cooling of gas in the centers of our subhalos could help to stabilize
them against disruption.  Our numerical resolution is high enough
that the relevant subhalos are not lost to numerical disruption in any case,
and our satellite fractions are at or above observational estimates
\citep[see][for a compilation]{WetWhi10}.
The outer envelope which is lost to stripping is expected to be mostly dark
matter, so this physics will be correctly modeled.  Once a majority of the
mass is lost the subhalo mass will be much less than the host halo mass, and
the amount of dynamical friction experienced will be small, mitigating any
error in the precise amount \citep{Sim09,LauNagKra10,JiaJinLin10}.
Extending high dynamic range simulations such as ours with additional physics which is in accord with observational constraints would be very interesting.

\section{Mock observations} \label{sec:mock}

Given the matter and subhalo distribution, we compute a number of mock
observations to investigate how the complicated nature of structure
formation influences observational probes of clusters.  In all cases we
use constant time outputs from the simulation, and consider the box
in isolation, i.e.~we do not attempt to make light cones, remap or stack boxes.
Our simulations contain sufficient path length to answer the questions
of interest to us here without needing to employ these techniques.
Also, we do not model the cluster finding process itself.  Rather
we ask about the measurements that could be made once a cluster was
correctly identified.

To identify correlations due to the anisotropic nature of the cluster
and its environment, we observe each cluster along 96 different lines of
sight, centering it within the periodic box.  (For intrinsic measurements in
\S\ref{sec:intrinsic} more sightlines are considered, when needed, as
described therein.)  
Each line of sight then is used to find galaxy richness, velocity
dispersion, lensing and integrated Compton distortion.
Our resulting sample has 83 clusters with
$M_{180b}\geq 2\times 10^{14}\,h^{-1}M_\odot$
along almost $8,000$ lines of sight total, and 242 clusters with
$M_{180b}\geq 10^{14}\,h^{-1}M_\odot$ along $\sim 23,000$ lines of
sight.

\subsection{Richness} \label{sec:rich}

The easiest property of a cluster to observe is its ``richness'', or the
number of galaxies it contains.  Each halo above any infall mass threshold,
$M_{\rm min}$, hosts one central subhalo above the same threshold mass,
and a number of satellite subhalos which is (approximately) Poisson distributed
about a mean $(M_{\rm halo}/M_1)$ with $M_1\sim 15\,M_{\rm min}$.
Unfortunately this information is not observationally accessible, and proxies
must be used.
There are numerous definitions of richness in the literature, here we
consider only two as representative of the class.

The first is the richness defined by \citet{Yan08}, which computes a
phase space density for each cluster and assigns galaxies above a
threshold to a cluster candidate.  The richness is the number of
galaxies assigned.  Rather than iterate our fit, we use the cluster's
true mass in the model, but otherwise implement the method as
they describe, including all galaxies within the simulation, not just true 
cluster members, in the calculation\footnote{We use $H/c$ rather than $H_0/c$ as
the prefactor in their equation 7.}.
As shown in Fig.~\ref{fig:yang} the richness measured in our simulations is in
quite good agreement with that inferred from the observations and the richness
does show strong trends with host halo mass.
However it does require knowledge of the spectroscopic redshifts of all
galaxies.  We call this quantity phase space richness below.

A second richness definition counts only those galaxies within the red
sequence and within an aperture, subtracting an estimate of the contamination.
The hope is that using only these galaxies reduces the impact of interloper
galaxies from large line-of-sight distance and blue galaxies in front of the
cluster, without requiring spectroscopic redshift information\citep[see, e.g.][] {RCSI,RCSII}.
This requires us to assign a color to each of our mock galaxies.
By abundance matching we are able to assign a luminosity (or stellar mass)
to all of our subhalos.  We use the method of \citet{SkiShe09} to further
assign them a color and we include them in the projected red sequence based
upon their distance from the true cluster redshift (this method has only
been calibrated at $z\simeq 0.1$ so we do not assign colors when considering
higher $z$).
Further details are given in Appendix \ref{app:color}.
The richness includes galaxies brighter than $0.4\,L_\star$ (i.e.~$M_\star+1$)
with the background subtraction computed precisely using the periodic
simulation volume.  The transverse aperture is set following the convention
used in the maxBCG catalog (\citealt{maxBCG}; see also \citealt{Hig10}):
a first estimate of the richness is obtained within a $1\,h^{-1}$Mpc transverse
radius and this richness is used to estimate $R_{200b}$ which is then used for
a final richness estimate.
Our final richness-mass relation (not shown) is in good agreement with the
scaling relation found in observed clusters.
Note that this procedure has the unwelcome property of increasing scatter due
to filament-based projection effects.  Should the initial estimate be high due
to projected galaxies the aperture will be set too large and include even more
galaxies.  We noted a large increase in richness scatter at fixed mass using
this procedure relative to when we use the true radii.

\subsection{Galaxy kinematics}

Modeling of galaxy kinematics in clusters remains a major tool in
determining their properties.  Since we are able to resolve and track
the subhalos which would host galaxies within our simulation, we are
in a good position to study how their velocity structure depends upon
and correlates with cluster properties and the larger environment.
As this capability is new in terms of mock velocity observations,
we shall develop it in some detail in the next two sections.
The intrinsic properties of the velocity field, including velocity bias,
were discussed in \S\ref{sec:galaxies}.
Anisotropy and substructure are discussed in \S\ref{sec:intrinsic}.
Our modeling of interloper rejection and dispersion estimation is
the subject of \S\ref{sec:interlopers}.

\subsection{Lensing}

The distribution of mass can be probed by the distortion of background
galaxy shapes due to the gravitational deflection of light by the
potentials of massive halos.  This signal is sensitive to the projected
mass along the line-of-sight, $\Sigma$, weighted by a kernel
\citep[e.g.][for recent reviews]{Hoe02,Ref03}.  The lensing kernel varies 
only very slowly with distance, so all of the matter in and around the cluster 
receives similar weight.  If we assume the source and lens redshift (distributions) 
are known, lensing measures the projected mass.  We do not attempt to 
model the full light cone here, rather we make the approximation that mass 
far from the cluster is uncorrelated with the cluster and contributes only 
a ``noise'' while mass close to the cluster receives the same weight as the cluster itself.
We ignore the noise term (and any additional noise from the finite number
of source galaxies or observational non-idealities) and approximate a
lensing observation as a measurement of the projected mass, apodized with
a Welch kernel
\begin{equation}
  W(z) = 1 - \left( \frac{2Z}{L_{\rm box}} \right)^2
\label{eqn:apodize}
\end{equation}
where $-\frac{1}{2} L_{\rm box} < Z < \frac{1}{2} L_{\rm box}$
is the line-of-sight coordinate.
The window vanishes for $|Z|>\frac{1}{2} L_{\rm box}$.
We model all lensing observations as applying to $\Sigma(R)$
determined in this manner along any line-of-sight, using
the periodicity of the box to place the lensed object at the center of
the box.

A detailed study of lensing projection effects is not the focus of this
paper.   It has been discussed in detail previously
\citep{RebBar99,MetWhiLok01,Hoe01,PutWhi05,Men10}. 
In order to gauge the approximate size of the effect and its degree of
correlation with other measures of cluster size we simply fit a
singular-isothermal-sphere model ($\rho\propto r^{-2}$) to the lensing
signal.  In order to remove much of the uncorrelated signal we use the
$\zeta$ statistic
\begin{equation}
  \zeta(R_0;R_1,R_2) \propto \left\langle \Sigma(R<R_0)\right\rangle -
                     \left\langle \Sigma(R_1<R<R_2)\right\rangle
\end{equation}
where the constant of proportionality depends on the source and lens
redshift distributions which we shall assume known for simplicity.
For our singular-isothermal-sphere this gives
\begin{equation}
  \zeta \propto \frac{\sigma_{\rm lens}^2}{G}
  \left( \frac{1}{R_0}-\frac{1}{R_1+R_2} \right)
\end{equation}
as a function of $R_0$ at fixed $R_1$ and $R_2$ which we use to fit for
$\sigma_{\rm lens}$.  The results are quite stable to variations in the $R_i$,
for our fiducial results we fit $R_0$ in the range $0.1\,r_{180b}$ to
$r_{180b}$ with $R_1=r_{180b}$ and $R_2=1.25\,r_{180b}$.
Qualitatively similar results are obtained if we fit directly to the
projected mass, $\Sigma(R)$, or use a different profile such as the broken
power-law of \citet{NFW}.

\subsection{Sunyaev-Zel'dovich effect}

Another method for finding and weighing galaxy clusters is to study the
distortion they introduce in the cosmic microwave background (CMB).  The
``hot'' electrons in the intra-cluster medium can scatter the ``cold'' CMB
photons to higher energy, distorting the spectrum in predictable ways
\citep{SunZel72}.  The surface brightness of the Compton distortion is
independent of distance, and the integrated signal is proportional to
the total thermal energy of the gas, making this a powerful means for
finding and characterizing clusters.
The insensitivity to distance, however, means that SZ experiments must
also contend with projection effects.  Assuming a self-similar cluster,
the Compton distortion scales as $M^{5/3}$, so lower mass halos contribute
fractionally less than they do to a lensing or galaxy measure, but the
relatively lower resolution of the observations exacerbates the problem.

In earlier work \citep{CohWhi09} we investigated optical and SZ methods for
finding clusters and found that the scatter from the cluster candidates in these two methods was correlated.
We continue that investigation here, using a simple model of the Compton
distortion appropriate to low-resolution observations (such as provided by
the South Pole Telescope\footnote{http://pole.uchicago.edu} or
Atacama Cosmology Telescope\footnote{http://www.physics.princeton.edu/act}).
We assign to each dark matter particle in the simulation a ``mean''
temperature based on the velocity dispersion of its parent halo, and
compute the total Compton distortion as a sum of the mass times the
temperature in cylinders, apodizing the signal as above (Eq.~\ref{eqn:apodize}).
This misses contributions to the temperature from e.g.~shocks, small-scale
structure in the intra-cluster gas and the run of temperature with radius.
For partially resolved cluster observations however the low-order properties
of the maps so obtained are in reasonable agreement with hydrodynamic
simulations which include these effects \citep[e.g.][]{WhiHerSpr02},
and serve to illustrate our main points.
We shall use as our observable the integrated Compton $y$-parameter within
a disk of radius $r_{180b}$, as this is a more stable quantity than
e.g.~central decrement.

\section{Halo Intrinsic properties} \label{sec:intrinsic}

\subsection{Halos}

\begin{figure}
\begin{center}
\resizebox{3in}{!}{\includegraphics{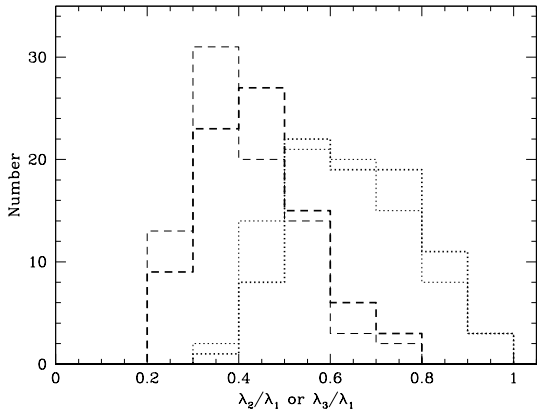}}
\end{center}
\caption{The distribution of $\lambda_2/\lambda_1$ (dotted) and
$\lambda_3/\lambda_1$ (dashed) for all halos more massive than
$2\times 10^{14}\,h^{-1}M_\odot$ at $z=0.1$.  Thick lines show the results for
subhalos within the FoF halos and thin lines for subhalos within $r_{180b}$ of
the most bound particle in each group.
Note that the eigenvalues, $\lambda$, of the velocity anisotropy tensor
scale as $\lambda\sim\sigma^2$.}
\label{fig:lamdist}
\end{figure}

We begin by considering the intrinsic properties of our massive halos and
their subhalo populations, absent any line-of-sight projection or
misidentifications.  It is well known that the 3D density profile of massive
halos is triaxial \citep{ThoCou92,War92,JinSut02},  with the major axis
approximately twice as long as the minor axes which are approximately equal
in size.  When spherically averaged the density profiles of the `relaxed' halos
resemble a broken power-law \citep{NFW} with the inner regions forming early
and then remaining approximately constant as subsequently accreted dark matter
is kept away from the center by the angular momentum barrier.
Our subhalos follow a profile similar to that of the dark matter, though
shallower in the central regions.

\subsection{Velocity ellipsoid} \label{sec:velellip}

\begin{figure}
\begin{center}
\resizebox{3in}{!}{\includegraphics{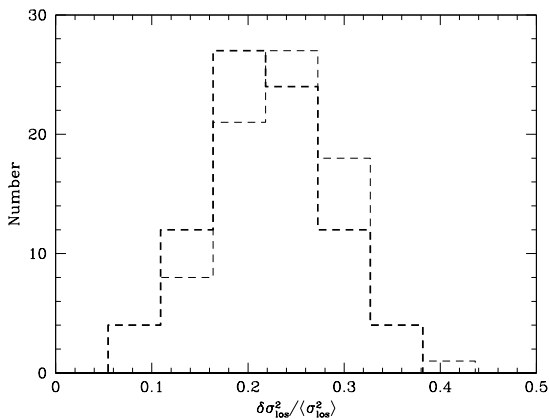}}
\end{center}
\caption{The distribution of
$\delta\sigma_{\rm los}^2/\langle\sigma_{\rm los}^2\rangle$
for all halos more massive than $2\times 10^{14}\,h^{-1}M_\odot$ at $z=0$.
As in Fig.~\protect\ref{fig:lamdist} thick lines show the results for
subhalos within the FoF halos and thin lines for subhalos within $r_{200}$ of
the most bound particle in each group.}
\label{fig:sigdist}
\end{figure}

Although the 3D, dark matter velocity dispersion within $r_{200c}$ is well
correlated with $M_{200c}$ \cite[e.g.][]{Evr08} and the galaxies show little
velocity bias compared to the dark matter, the line-of-sight velocity
dispersions show considerably more scatter.
The galaxy line-of-sight dispersion for any viewing angle, $\hat{n}$, is simply
$\sigma_{\rm los}^2=\hat{n}^T\cdot\sigma\cdot\hat{n}$, where $\sigma^2_{ij}$ is
the anisotropy tensor,
$\sigma^2_{ij}=\langle (v-\bar{v})_i(v-\bar{v})_j\rangle$,
averaged over subhalos in the host halo.   
As has been noted before in the dark matter particles \citep{Tor97,KasEvr05}, and seen
here for the galaxy subhalos as well, the velocity tensor, like
the moment of inertia tensor, is quite anisotropic
(see also Fig.~\ref{fig:dotplot}).
Not surprisingly the principal axes of the two are quite well aligned,
with a typical mis-alignment angle of $\sim 20-30^\circ$.

If we order the eigenvalues of $\sigma_{ij}^2$ as
$\lambda_1>\lambda_2>\lambda_3$
then for uniformly chosen $\hat{n}$ the distribution of $\sigma_{\rm los}^2$
has a peak at $\lambda_2$, a mean at
$\frac{1}{3}(\lambda_1+\lambda_2+\lambda_3)$
and a width
\begin{equation}
  \left(\delta\sigma_{\rm los}^2\right)^2 = \frac{4}{45}
  \left[ \lambda_1^2+\lambda_2^2+\lambda_3^2
    - \lambda_1\lambda_2 - \lambda_2\lambda_3 - \lambda_3\lambda_1 \right]
\label{eq:sigscatter}
\end{equation}

For our sample,
the distribution of eigenvalues for all halos above
$2\times 10^{14}\,h^{-1}M_\odot$ at $z=0$ is shown in
Fig.~\ref{fig:lamdist}, where we see typical values for
$\lambda_3/\lambda_1$ and $\lambda_2/\lambda_1$ are $0.3$ and $0.6$
respectively.
For the more massive subhalos the spread in eigenvalues is slightly larger
than for a random subset of the mass but they become increasingly comparable
as we move down the subhalo mass function.

We found that the distribution of measured velocity dispersion
for any cluster, along 10,000 randomly selected lines of sight, tended to be
significantly non-Gaussian.
The distribution of $\delta \sigma_{\rm los}^2$ from 
Eq.~\ref{eq:sigscatter} is shown for our massive halos in Fig.~\ref{fig:sigdist}, 
we see that
$\delta\sigma_{\rm los}^2/\sigma_{\rm los}^2$ is peaked at 20-30 per cent.
If one assumes
$M\propto\sigma_{\rm los}^3$ this gives an inferred mass error of nearly 40
per cent.
This suggests that,
even absent any interlopers, velocity bias, or observational non-idealities,
velocity dispersion mass estimators will work better in an ensemble sense than
for any individual cluster.   

\begin{figure}
\begin{center}
\resizebox{3in}{!}{\includegraphics{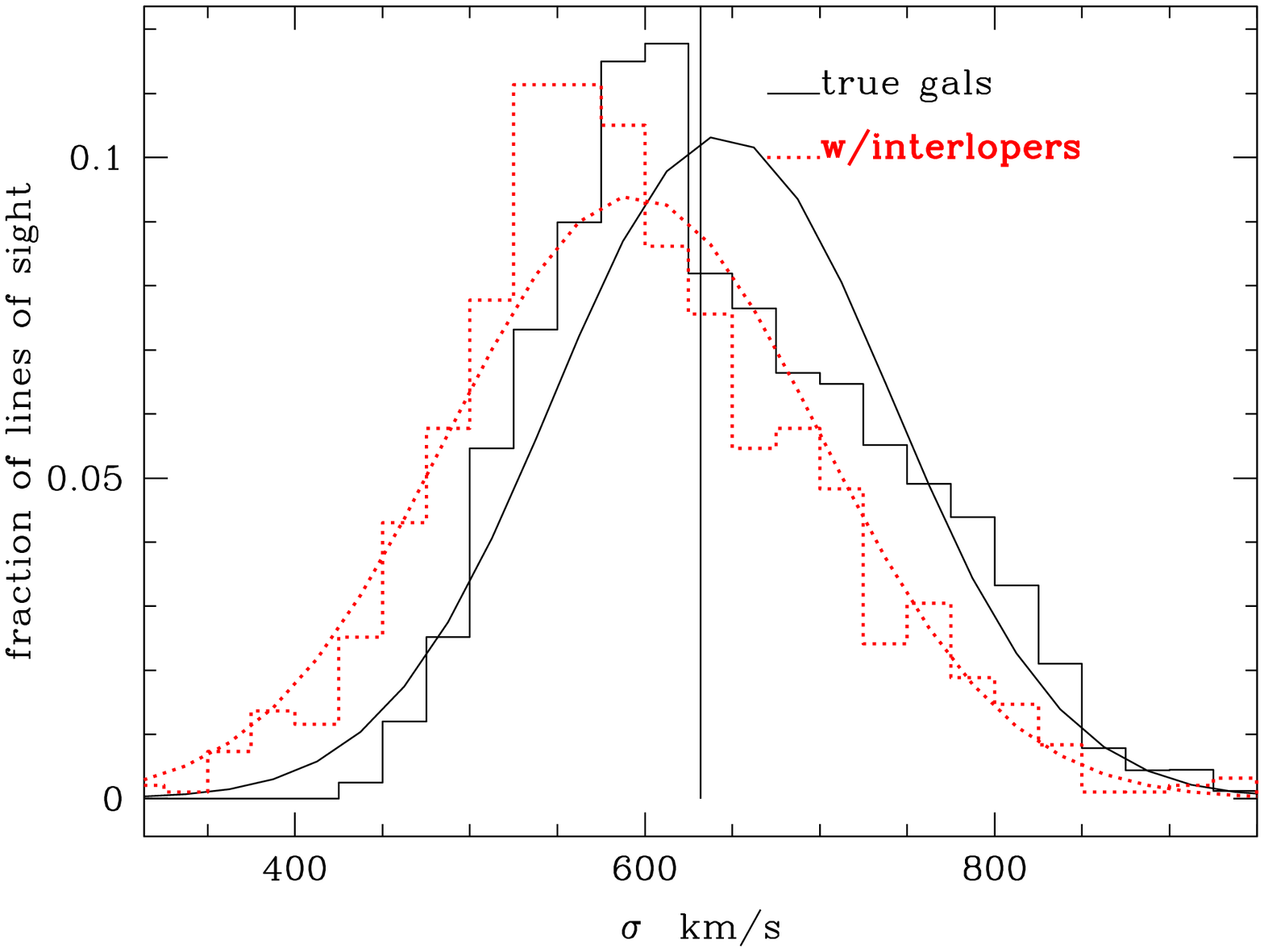}}
\end{center}
\caption{The composite histogram, normalized to unit area, of velocity
dispersions for 10 clusters with
$3\times 10^{14}\,h^{-1}M_\odot\leq M_{180b}\leq3.5\times 10^{14}\,h^{-1}M_\odot$,
in two cases: with only the true group members included
(solid, black line, with 10,000 lines of sight per cluster)
and with the full line-of-sight with interlopers removed as in
\S\protect\ref{sec:interlopers}
(red, dotted line, with 96 lines of sight per cluster).
The curves indicate Gaussian fits.
The vertical line at $\sigma\simeq 620\,{\rm km}/{\rm s}$ is the average
of the three-dimensional velocity dispersions for these clusters.
The line-of-sight measurement with interloper rejection has a lower average
velocity dispersion \protect\citep[as seen also in][]{vHaarlem97,Biv06}.
Even this narrow range of halo masses exhibits a wide range of $\sigma$,
as discussed in the text, with Gaussian fits of width
$100\,{\rm km}/{\rm s}$.
The skewness in the distribution is not present for all samples of this size.}
\label{fig:velbiv}
\end{figure}

As an example of an ensemble measurement, Fig.~\ref{fig:velbiv} shows the
distribution of velocity dispersions measured from our simulation for 10
clusters with
$3\times 10^{14}\,h^{-1}M_\odot\leq M_{180b}\leq 3.5\times 10^{14}\,h^{-1}M_\odot$.
The solid histogram is the composite of $\sigma$ values for all the clusters,
using the member galaxies only and projecting along 10,000 lines of sight
for each cluster, while the line shows a Gaussian fit.
(The dotted line and histogram are for the distribution which results when
the same clusters are observed along 96 lines of sight, including interlopers
and a culling method discussed below in \S \ref{sec:interlopers}).

\begin{figure}
\begin{center}
\resizebox{3in}{!}{\includegraphics{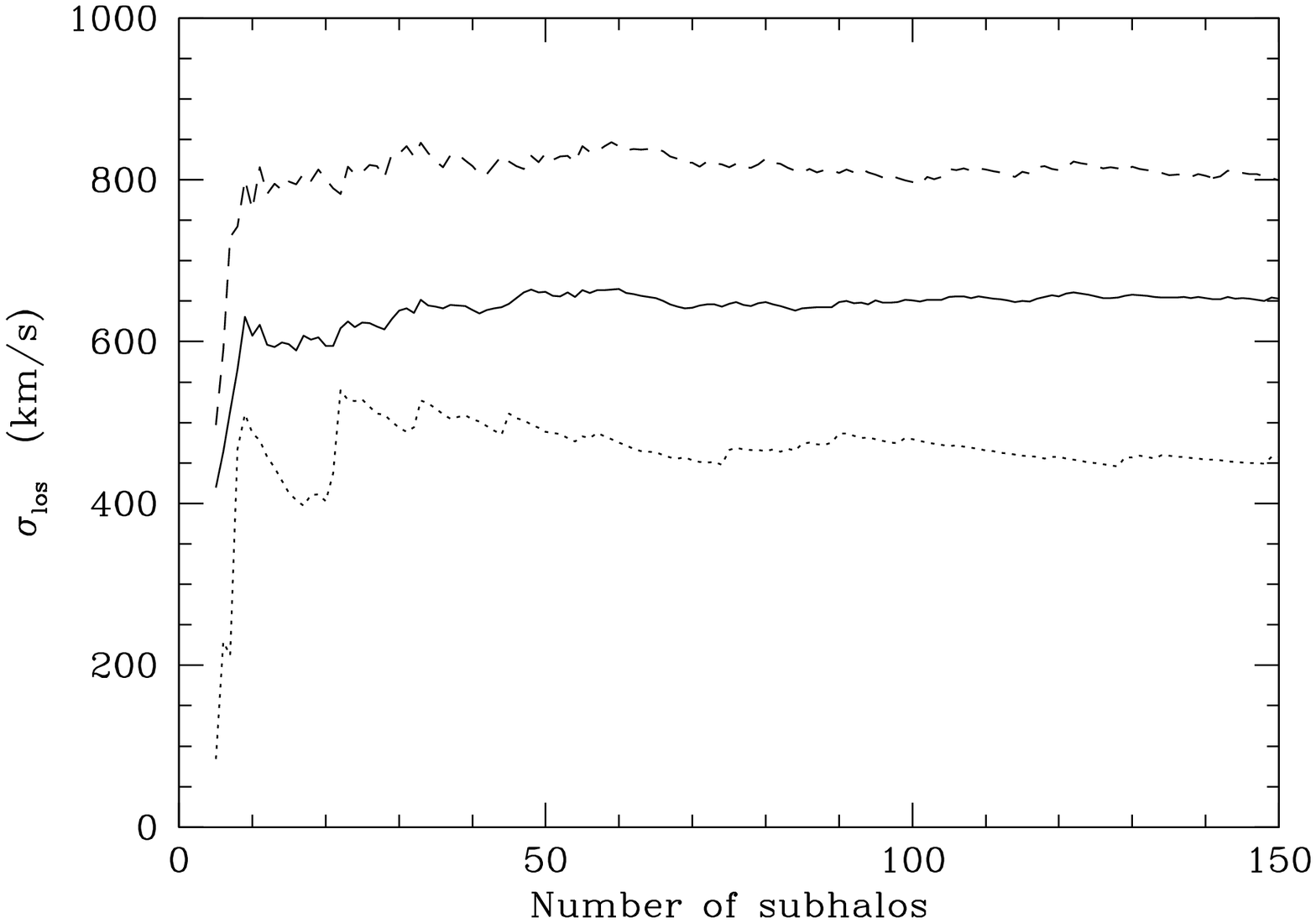}}
\end{center}
\caption{The line-of-sight velocity dispersion vs.~the number of subhalos
included (ordered by $M_{\rm inf}$, i.e.~luminosity, from highest to lowest)
for a halo with $z\simeq 0.1$ mass $4\times 10^{14}\,h^{-1}M_\odot$.
We include only subhalos which lie within the friends-of-friends halo,
excluding any interlopers.
The solid line shows the isotropic dispersion (i.e.~$\sigma_{3D}/\sqrt{3}$)
while the dotted and dashed lines show the dispersion along the eigenvector
directions corresponding to the smallest and largest eigenvalues of
$\sigma^2_{ij}$.}
\label{fig:sig_vs_nsub}
\end{figure}

\begin{figure}
\begin{center}
\resizebox{3in}{!}{\includegraphics{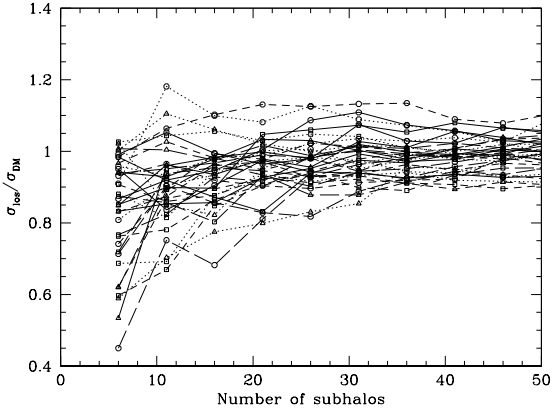}}
\end{center}
\caption{As in Fig.~\protect\ref{fig:sig_vs_nsub}, the line-of-sight velocity
dispersion vs.~the number of subhalos included (ordered by $M_{\rm inf}$ from
highest to lowest), but now for a range of halos and focusing on the region
$N_{\rm sub}\le 50$.  As before, we include only subhalos which lie within
the friends-of-friends halo, excluding any interlopers, and plot the
line-of-sight dispersion normalized by the dispersion of the dark matter
within the friends-of-friends halo.  Note that each line converges to a
stable result for large numbers of subhalos but the value depends upon the
cluster and line-of-sight chosen, as does whether the approach is from above
or below.}
\label{fig:sig_vs_n}
\end{figure}

This intrinsic line of sight scatter also
suggests that if the goal is to determine the mass distribution there
is an upper limit to the number of galaxy redshifts per cluster it is desirable to
obtain: there is little to be gained by reducing sources
of error in $\sigma_{\rm los}$ significantly below the dispersion above.
Figure \ref{fig:sig_vs_nsub}, which shows the velocity dispersion as a function
of the number of subhalos used (added in order of decreasing luminosity),  
gives an illustration of this.
Only subhalos which are within the friends-of-friends halo are included.
All of the measures converge to a stable value for large numbers of subhalos,
but the value depends significantly on the chosen line-of-sight.
We find the number of subhalos at which the asymptotic limit is reached,
and whether that approach is from above or below, depends upon the cluster
under consideration, but the results are generally stable once 50 subhalos
are included (Fig.~\ref{fig:sig_vs_n}).


\begin{figure*}
\begin{center}
\resizebox{6in}{!}{\includegraphics{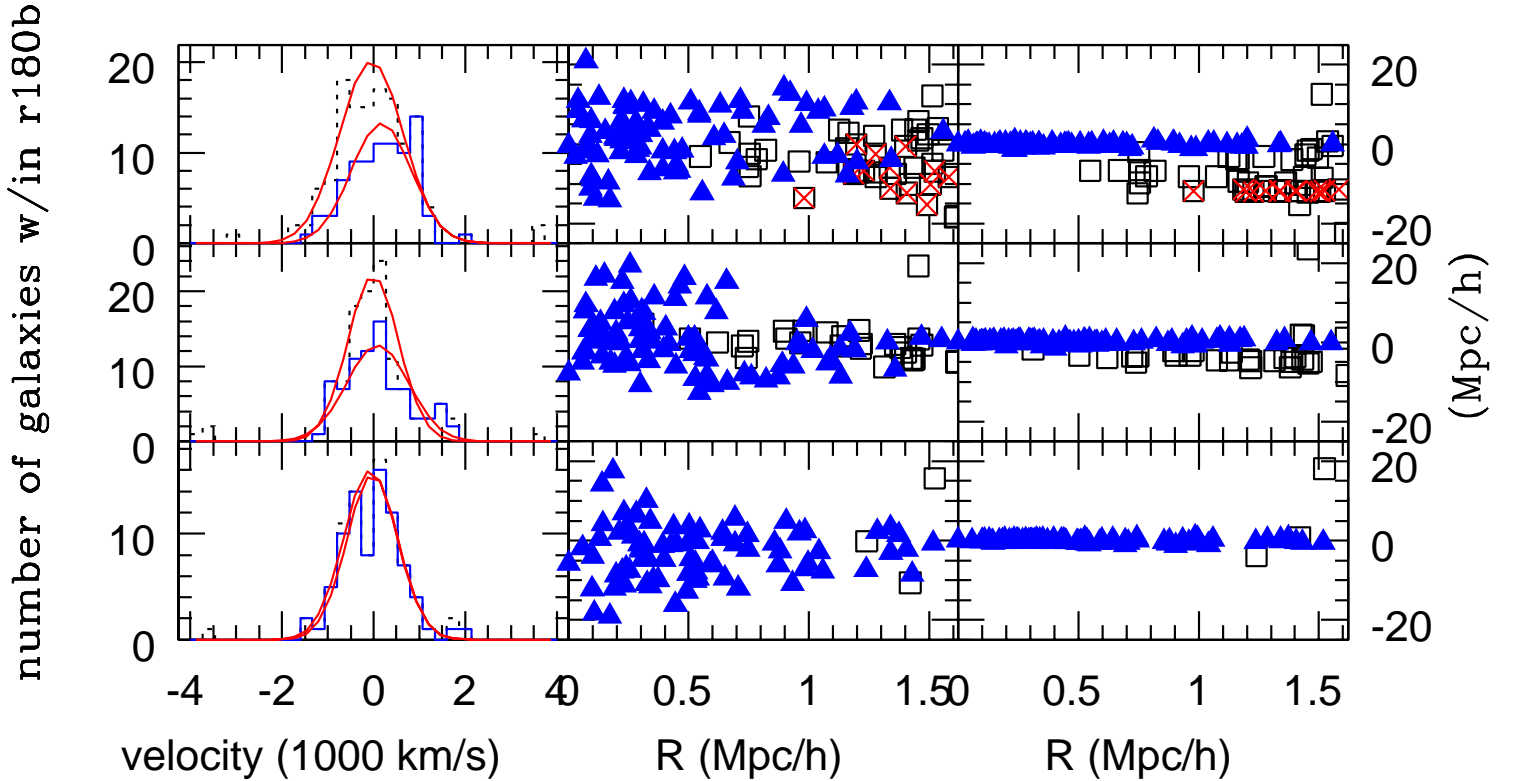}}
\end{center}
\caption{Line of sight histograms for the same massive
$(M_{180b}=2.4\times 10^{14}\,h^{-1}M_\odot)$ cluster along three
different lines of sight.
(Left) velocity histograms for all galaxies within $r_{180b}$ in plane of sky
(dotted) and true cluster members (solid).  The smooth curves are Gaussians
with the same area under the curve as the line of sight velocity distributions,
and with dispersion fit to the core of the line of sight velocity distribution
to guide the eye. 
(Center) observed phase space diagram, transverse radius vs.~redshift space
position (including peculiar velocities).
Solid (blue) triangles are true cluster members, open squares are interlopers
{}from halos with mass $<0.2\,M_{\rm clus}$ and squares with (red) crosses
inside are interlopers from halos with mass $>0.2\,M_{\rm clus}$,
i.e.~massive neighbors.
(Right) true phase space diagram, transverse radius vs.~true line-of-sight
position (absent peculiar velocities).
The bottom row is for a sightline where there are only 3 nearby galaxies
(out of 87) and all measures are within 50 per cent of the true mass,
the middle row has many nearby galaxies, but none from
massive halos and overpredicts the mass in both Compton distortion and weak
lensing by at least 50 per cent, and the top row has nearby galaxies from a
nearby massive halo, about $10\,h^{-1}$Mpc in the foreground, and overpredicts
the mass from lensing, velocity dispersions and Compton distortion.
None of these lines of sight have appreciable substructure using the
Dressler-Shechtman test described in \S\protect\ref{sec:substructure}.}
\label{fig:hist3} 
\end{figure*}

Fig.~\ref{fig:hist3} shows some typical line-of-sight velocity histograms
and phase-space distributions for a massive
($M_{180b}=2.4\times 10^{14}\,h^{-1}M_\odot$)
cluster viewed down three different lines-of-sight, with the solid lines
being the histogram for the galaxies found within $r_{180b}$, i.e.~the
``true'' cluster members.  There is a large variation in the velocity
dispersion profiles, even when only true members are included.
The interloper structure seen will be discussed in \S \ref{sec:interlopers}.

\subsection{Substructure} \label{sec:substructure}

\begin{figure*}
\begin{center}
\resizebox{7in}{!}{\includegraphics{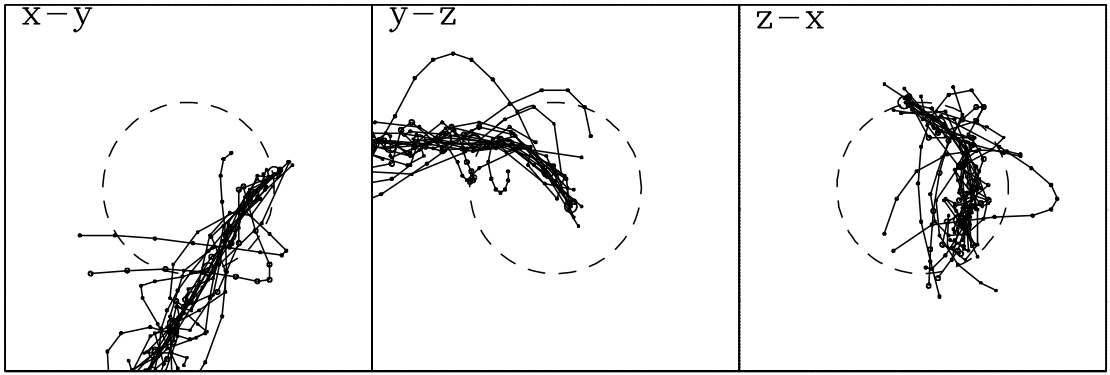}}
\end{center}
\caption{The persistence of substructure.  The tracks show a small subset of
the subhalos which fell into a large halo in the simulation as part of a large
group at $z\simeq 0.3$, corresponding to the last major merger for this halo.
Each panel is $6\,h^{-1}$Mpc on a side, centered on the $z=0$ position of
the most bound particle in the halo, and the dashed line marks the virial
radius ($r_{200c}$) of the main halo.
To avoid crowding only a small fraction of the subhalos, the main
progenitors with $M_{\rm inf}>10^{12}\,h^{-1}M_\odot$, are plotted.
Subhalos which merge with these halos before $z=0$, and any subhalos which
were not part of this group back at $z\simeq 0.3$ are omitted.  Note the
coherence of this ``group of subhalos'' for $3\,h^{-1}$Gyr.}
\label{fig:track}
\end{figure*}

Our massive halos contain significant substructure in both physical and
velocity space, which is frequently attributed to the active merger
histories of massive halos.
We find that groups of subhalos which fall in together remain highly
correlated for significant spans of time (several Gyr).
In many respects these past accretion or merger events are still ``ongoing'',
in that the 3D density field has multiple distinct maxima and one can still
see kinematically distinct groups of subhalos which were part of the merger
partner and fell in together at that time.
One example is given in Fig.~\ref{fig:track}, which shows the tracks of a
small subset of the subhalos in a massive cluster and illustrates the long-term
coherence of the group of subhalos even as it moves within the virial radius
of its host.
Though they are not highlighted in the figure, there are several other major
groupings of subhalos that were accreted together and have survived for some
time.  Each has had a complex merger history but shows a long term persistence
even though it is now well inside the formal virial radius of the host halo.  
It is an over-simplification to assume that when a halo falls into a larger
neighbor and becomes a satellite that all its satellites become associated
with the larger halo and evolve independently.

Not only do halos survive as distinct entities but groups of subhalos
do as well.  In fact, for massive halos, 30\% of the subhalos in our sample
are satellites when they fall in.
One consequence of this has been seen in other contexts: satellites often
merge with other satellites (rather than the central galaxy of their current
halo), and the satellite they merge with is often the old central of the halo
they were in prior to the merger \citep{Angetal09,Sim09,WetCohWhi09}.
Visually we also saw correlated velocities between nearby satellites with
different originating groups, presumably due to infall along a common filament.
This long-term dynamical coherence also indicates that care should be taken
when assuming relative velocities between galaxies are a substantial fraction
of the host virial velocity, e.g.~when estimating merger rates or impulses.

All of our simulated clusters have very obvious substructure.
We have implemented several standard tests for dynamical substructure, which
have been frequently applied to simulations and observed clusters in the
literature, to see how well they find the substructure we know to be there.
An excellent review of these methods can be found in \citet{Pin96};
some more recent statistics are summarized in \citet{Hou09}.
We focus on the three dimensional test of \citet{DreShe88},
and the one dimensional tests of Kolmogorov and Arnold-Darling described
in \citet{Hou09} and refer the reader to papers there for details.
The \citet{DreShe88} test has been applied to simulations previously
\citep[e.g.][]{Cen97,KneMul00}, but usually to a large subset of the dark
matter particles in the cluster rather than subhalos.  Because we use subhalos
identified with galaxies within the simulation, our methods are a further step
in quantifying the difficulty of identifying cluster substructures
observationally.
When dark matter particles are used the large number available allows them
to trace the cluster structure more faithfully than the observationally
available galaxies, but using random dark matter particles as sample
galaxies misses the dynamical coherence of groups of substructures that
naturally arises in hierarchical structure formation scenarios.

We find many of our clusters show signs of substructure along some lines of
sight: surprisingly, we find none of the three substructure indicators is well
correlated with the time since last major merger, and the values of the
indicators are very dependent on viewing angle for a given cluster even
before we consider interlopers due to projection\footnote{This is in contrast
to \citet{Cen97} who only found a large amount of substructure after interlopers
were included; one possible source of the difference is our use of galaxy
subhalos rather than dark matter particles.  Our results lend support to
\citet{CroEvrRic96} who found such tests perform relatively poorly as
cosmological indicators.}.
If the substructure is well separated along the line-of-sight, from the
bulk of the galaxies, then it is caught by each of the indicators.
Otherwise it can be missed.
As an example we pick one cluster, containing 57 subhalos brighter than
$M_\star+1$.	
When viewed down the $z$-axis it is not flagged as having substructure
by the tests we consider: the probability-to-exceed for the
Dressler-Shectman test is 54 per cent, $D^\star=1.18$ and $A^{2\star}=1.68$
(\citet{Hou09} suggest that $D^\star$ in excess of 1.2 or $A^{2\star}$ in
excess of 1.9 indicate the presence of substructure).
However, viewing this same cluster down the $x$-axis the probability-to-exceed
for Dressler-Shectman is 3 per cent, $D^\star=2.76$ and $A^{2\star}=10$.
There are many similar examples.
In some cases the Dressler-Schechtman test flags substructure where the
other tests do not, while for others the situation is reversed.
Sometimes the Dressler-Shectman $\Delta$ statistic is high, but similar or
larger values are obtained when shuffling the velocities
\citep[as described in][]{DreShe88} leading to a higher probability-to-exceed
or a lower significance detection of substructure.
In these cases the prevalence of substructures in the host halo means that
the ``shuffled'' statistics are not faithfully representing the
``no substructure'' scenario, leading one to erroneously assume the observed
value of the statistic is consistent with no substructure.

These results suggest caution when interpreting lack of observed
substructure in the galaxy distribution as evidence for a dynamically
relaxed, steady-state object (e.g.~justifying the use of the virial theorem
or Jeans analysis without the time derivative).
A cluster can be undergoing substantial mass accretion, i.e.~be far from
steady state, and still not be seen to have substructure along some lines
of sight.  
The viewing angle dependence also complicates inferences about incidence of
dynamical evolution of cluster galaxies from observed interactions of
subclusters within the cluster identified through substructure finding
techniques.
There are some indications \citep[e.g.][]{Biv96,Ada05} that more sophisticated
substructure finding techniques could yield more complete information in the
limit of hundreds of spectroscopic redshifts per cluster.  Since we found
earlier that the dynamics of the subhalos approached that of the dark matter
particles as we progressed down the subhalo mass function, we expect very
minor differences with earlier work when hundreds of subhalos are included.

\section{Interlopers} \label{sec:interlopers}

The intrinsic line-of sight scatter in velocity dispersion 
discussed above (\S\ref{sec:velellip}) is a
``best case'' estimate, where we have perfect identification of
cluster members.
In observations, an extra complication is provided by ``interloper''
galaxies which lie close to the cluster in the plane of the sky and in
velocity but which nevertheless are really members of a different halo.
Restricting samples to elliptical galaxies or matching on photometric
properties can help, but does not solve the problem completely.
Conversely, measurement errors in the velocities (which we do not model)
can exacerbate the interloper problem -- though it is expected that
typical velocity errors will have only a small effect on estimates
\citep{Biv06}.

Returning to Fig.~\ref{fig:hist3}, we now turn attention to the interlopers
in the line-of-sight velocity histograms.  In the middle and right columns
we see the galaxies in phase space and physical space with true members
represented by filled triangles, interlopers represented by open boxes
and interlopers from massive halos (with mass $>0.2\,M_{\rm clus}$)
represented by open boxes with crosses in them.
Depending upon the line of sight, the same cluster can have (top to bottom):
contributions from nearby massive halos, contributions only from less massive
halos, or few interlopers.
In these three instances the inferred red-sequence richness is very high,
high and close to the mean for a cluster of this mass.
It is typical that a single halo exhibits each of these characteristics
when viewed from different directions, as a large fraction of halos have
a massive neighbor.
For the ensemble of velocity dispersions in our sample, while the
distributions can often be well fit by a Gaussian profile, a non-trivial
fraction of the lines-of-sight lead to ``flat topped'', skew or bi-modal
distributions or distributions that can be fit with Gaussians plus an excess
in the wings \citep[as seen in observations, e.g.][]{Mil08}\footnote{Stacking
the velocity histograms for all lines of sight corresponding to some richness
or mass range also yields an approximate Gaussian, with excess in the far wings,
which also has been observed \citep[e.g.][]{vdMarel}.}.
In some cases interlopers cause an excess in the center of the velocity
distribution.

\subsection{Interloper removal}

Several techniques have been devised to identify and reject these interlopers.
Since in the simulations we know which objects are true
cluster members we can apply these algorithms to our samples to see
how they perform.
Such investigations have been done before
\citep[e.g.][]{Per90,denHartog,vHaarlem97,Cen97,Dia99,Lokas06,Woj07,Woj09}
but typically using randomly selected dark matter particles rather than
subhalos.
By using subhalos we keep any correlations between subhalo positions and
large-scale environment or between subhalos which fell in together as part
of a larger structure.   Including the interlopers in
our mock observations and then using observational techniques to 
attempt to remove them is also important for estimating the
scatter induced by the cosmic web, a main concern of this paper.

One of the simplest, and most widely used, interloper rejection methods is
$3\,\sigma$ clipping (\citealt{Yahil1977,Lokas06}; it has been applied in some
large surveys and individual objects of special interest such as
\citealt{Hal04,GalLub04,Bec07,Mil08,Kur09}),
which uses the fact that line-of-sight velocities of cluster members are
close to Gaussian and iteratively excludes all galaxies $3\,\sigma$ away from
the mean.
Given enough galaxies one can perform this procedure in bins of transverse
radius.  \citet{Per90} developed a method based on removing galaxies whose
absence causes the largest change in a mass estimator while
\citet{DiaGel97} proposed the use of caustics and \citet{Pra03}
proposed an escape velocity cut.
Various authors argue that the `gaps' in the velocity distribution give
a better rejection criterion (\citealt{Zab90}, \citealt{Kat96},\citealt{Owers}).
Methods which use both projected coordinates and velocity
information were introduced by \citet{denHartog,Fadda96}.

We tested a number of interloper rejection algorithms.  Here we focus on
an example of the more complex methods which uses projected coordinates and
velocity information \citep[see e.g.][and references therein]{denHartog,Biv06,Woj07},
and its comparison to simple $3\,\sigma$ clipping.  
Such comparisons have been performed before \citep{vHaarlem97,Woj09} but
usually using randomly selected particles from lower resolution simulations
rather than subhalos.  Again this means that correlations between observed
galaxy properties are more faithfully tracked in our case. 

Our implementation is as follows.
We assume that a trial center of the cluster has been determined.
All galaxies with velocities within $3,000\,$km/s of the central velocity and
projected radius smaller than $r_{180b}$ are then selected.
The weighted gap method \citep[e.g.][]{Gir93} is then used to further remove
galaxies along the line-of-sight.  Specifically, gaps are defined as
$g_i=v_{i+1}-v_i$ for the sorted velocities and weights as $w_i=i(N-i)$,
for $i=1,\ldots,N-1$ for $N$ galaxies.  Galaxies to one side of a 
weighted gap larger than 3 are removed, where the
weighted gap is defined as 
\begin{equation}
\frac{\sqrt{g \cdot w}}
{MM(\sqrt{g \cdot w)}} \; ,
\end{equation}
with the midmean of the weighted gaps defined as
\begin{equation}
MM(\sqrt{g\cdot w})=\frac{2}{N}\sum_{N/4}^{3N/4}\sqrt{g\cdot w}
\quad .
\end{equation}
The motivation for such a cut lies in the expectation that the velocity
dispersion is Gaussian, and the assumption that when the galaxy distribution
departs from a Gaussian ``core'' it is no longer associated with the
halo of interest.
We use a modification to the weighted gap as described by \citet{Owers},
where it is applied separately in annuli of 50 galaxies each\footnote{The
results were stable for between 25 and 50 galaxies per annulus, for
definiteness we used 50.}.  Otherwise, we found the weighted gap tended
to throw out too many galaxies.

Now we use the projected distribution to define a further, transverse
radius dependent, velocity cut and iteratively remove galaxies beyond this
cut.  The cut depends on the (projected) harmonic radius, defined as
\begin{equation}
  R_h^{-1} \equiv \frac{2}{N(N-1)} \sum_{i<j} R_{ij}^{-1} \quad ,
\label{eqn:Rhdef}
\end{equation}
where the sum is over galaxies out to radius $R$.
If the velocity dispersion of the currently remaining galaxies is $\sigma$,
we define a circular velocity as
\begin{equation}
  v_c^2(R) = 3\pi\ \sigma(R)^2\ \frac{R_h(R)}{R} \quad .
\end{equation}
Typically $R_h/R\approx 1/2$ and decreases from center to edge, as the
profile becomes steeper\footnote{For example, if the density profile is a
power-law, $\Sigma(R)\propto R^{-p}$, the ratio $R_h/R=(3-2p)/(4-2p)$.}.
{}From $v_c$ we further define a ``freefall'' velocity, $v_{ff}=\sqrt{2}\,v_c$.
Then a galaxy at (projected) radius $R$ is an interloper if it is further from
the cluster center than
\begin{equation}
  v_{\rm max}(R) = {\rm max}\left[v_{ff}(R)\cos\theta,v_c(R)\sin\theta\right]
\end{equation}
where $\theta$ is the angle between radial vector and the line-of-sight
and the maximization is done over the (real-space) line-of-sight position
of the galaxy (the idea being that any observed galaxy can either be on a
circular or radial orbit, with different boundedness criteria).
Finally the velocity dispersion of the remaining galaxies is estimated using
the bi-weight estimator described in \citet{Beers}, and we shall refer to
this as $\sigma_{\rm kin}$.

Though there is a wide diversity in cluster behaviors, the method
of interloper rejection is more important than the precise dispersion estimator.
The use of the on-sky positions to define a transverse radius dependent
velocity cut performs slightly better than a fixed threshold, but in both
cases the threshold varies significantly from step to step and can remove
true cluster members while keeping actual interlopers.
\begin{figure}
\begin{center}
\resizebox{3in}{!}{\includegraphics{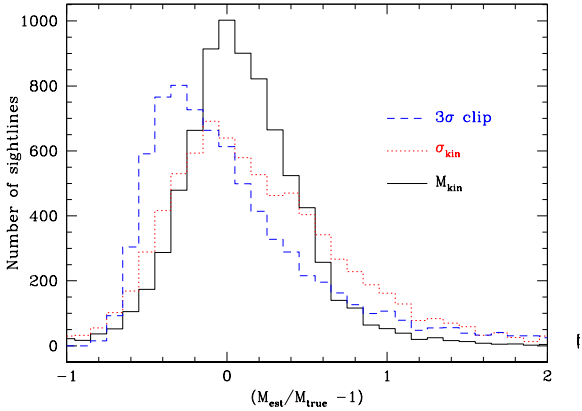}}
\end{center}
\caption{The distribution of mass predictions for the ensemble of sightlines
for our massive sample using $3\,\sigma$ clipping, $\sigma_{\rm kin}$ and
$M_{\rm kin}$ as described in the text.  The areas under the curves differ
because extreme outliers extend beyond the x-axis range shown.
The ``virial'' mass, $M_{\rm kin}$, is the best tracer of the true mass.}
\label{fig:sigcomp}
\end{figure}

Fig.~\ref{fig:sigcomp} compares results from $3\,\sigma$ clipping and our
more complex, phase-space based interloper rejection scheme for a sample
of massive clusters.  Except for extreme outliers, where the phase-space
method performs slightly better, the distributions are quite close and
noticeably non-Gaussian.
These results are quite insensitive to cluster mass.

The more complex algorithm can fail in some instances.  We found the most
sensitive step was the weighted gap measurement, which can fail when the
interloper structure along the line of sight is too close to define a clear
gap in the velocity histogram.
This is the case, for example, when two clusters are fairly close in one
line of sight or when we see a chain of small substructures close together,
as one would expect when looking down a filament. 
In these cases the weights given to the gaps do not work properly and
gaps are not properly detected.

\begin{figure}
\begin{center}
\resizebox{3.5in}{!}{\includegraphics{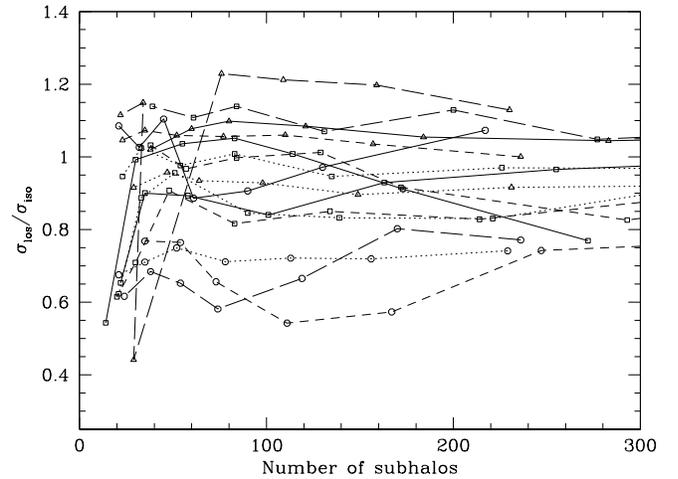}}
\end{center}
\caption{As in Figs.~\protect\ref{fig:sig_vs_nsub} and
\protect\ref{fig:sig_vs_n}, the line-of-sight velocity dispersion,
(in units of the isotropically averaged dispersion for all subhalos)
as a function of number of subhalos used, but now including non-member
subhalos and using the interloper rejection scheme described in the text.
The $x$-axis gives the number of subhalos used to compute the dispersion,
after interloper removal.  We have plotted 3 lines-of-sight for halos
with $M_{180b}$ in the range $(0.6-1.0)\times 10^{15}\,h^{-1}M_\odot$.}
\label{fig:changenbiv}
\end{figure}

Once we include line-of-sight projections and the need for interloper
removal, there is some gain to having more galaxies in order to better
estimate the cluster potential (Fig.~\ref{fig:changenbiv}) but the
intrinsic scatter due to the velocity ellipsoid remains a fundamental
limitation (e.g.~Fig.~\ref{fig:velbiv}).
As the number of galaxies with which we estimate the dispersion increases
the estimate becomes stable but is still a relatively poor estimate of the
angle-averaged dispersion.

\subsection{Degradation due to interlopers}

Applying this technique to our clusters, along 96 lines of sight each,
allows us to find and compare the distributions of $\sigma$ values
resulting from interlopers (and their rejection as described above) and
the distribution due to intrinsic line of sight variation.
The dotted line in Fig.~\ref{fig:velbiv} shows the distribution of
$\sigma_{\rm kin}$ for the 96 sightlines for 10 clusters in mass range
$(3-3.5)\times 10^{14}\,h^{-1} M_\odot$.
The standard deviation in $\sigma_{\rm kin}$ is about
$100\,{\rm km}/{\rm s}$ when only cluster member galaxies are included
and is approximately 10 per cent larger when including (and then rejecting)
interlopers\footnote{An study of cluster velocity dispersions \citep{Wei09}
applies the bi-weight estimator to a subset of clusters in the Millennium
simulation and also finds a large scatter (their Figure 1).}.
There is a slight downward shift in the mean $\sigma_{\rm kin}$, as was
also seen by \citet{Biv06}.
These trends are reproduced for higher and lower mass clusters.

The line-of-sight dispersion is only one piece of information available to
estimate masses, and other information can be introduced.  For example,
one can include the compactness of the cluster, estimated from the projected
member positions, or go further including corrections for surface terms and
orbital anisotropy, and beyond. 
It is not our intention here to model each of the (complex) methods which
have been presented in the literature, but we do note that the next-to-simplest
suggested mass estimator is proportional to $\sigma_{\rm kin}^2R_h$ where $R_h$
is the (projected) harmonic radius of Eq.~(\ref{eqn:Rhdef}).
We shall denote this $M_{\rm kin}$.
Formally this estimator would be valid only for spherical, isolated systems
with galaxies tracing mass, but due to a correlation between errors in $R_h$
and $\sigma$ we find $M_{\rm kin}$ produces a tighter, less skewed estimate
of $M_{180 b}$ than the pure dispersion based measures
(see Fig.~\ref{fig:sigcomp}) even in our more complex systems.
Although there is some variation from cluster to cluster, most often a
higher-than-average $\sigma$ is compensated by a lower-than-average $R_h$.
The compensation is not perfect, but it reduces the significance of the
fluctuation, leading to more lines-of-sight within the core of the distribution
and fewer strong outliers.
(Similar cancellations were seen by \citet{Biv06} when comparing observations
with and without interlopers.  They found the tendency of interlopers to bias
$\sigma$ low was (over-)compensated by their tendency to bias $R_h$ upwards.)
For this reason it is $M_{\rm kin}$ which we correlate with other quantities
in the following section.

We also note that \citet{Biv06} found a correlation between catastrophic
outliers in the mass-$\sigma_{\rm kin}$ or mass-$M_{\rm kin}$ relations
and substructure. 
Using the Dressler-Shectman test on the galaxies which were selected using
our interloper rejection procedure we found that 39 per cent of the lines
of sight had substructure ($P<0.05$) and of these only 10 per cent had $>50$
per cent deviations between $M_{\rm kin}$ and true mass.
By contrast, of the lines-of-sight with $>50$ per cent deviation in
$M_{\rm kin}$, 52 per cent had substructure to be compared to 40
per cent for lines-of-sight where $M_{\rm kin}$ is a reliable estimate of mass.
Thus outliers in $M_{\rm kin}$ do tend to have detectable substructure more
often than non-outliers, but substructure doesn't necessarily lead to
$M_{\rm kin}$ outliers and thus is not a reliable flag for it.

\section{Multiwavelength measures: correlated scatter} \label{sec:observed}

Although clusters obey tight scaling relations, we expect a large scatter
in individual measures of cluster mass/size.  Clusters are generally triaxial
and highly biased.  They are formed and fed at the intersection of a network
of filaments in atypical and anisotropic cosmological environments.
Their mass accretion is punctuated by a series of mergers with other massive
objects.
With the growing number of multi-wavelength, large area surveys underway
multiple measurements of large numbers of clusters are possible, and there
is a hope that different methods can cross-check each other.

As scatter is often caused by the cosmic web, measures sensitive to the
web will have correlations induced in their scatter.
Consider an idealized model, in which galaxies flow into the cluster from
a small number of (approximately straight) filaments, retaining memory of
this due to incomplete virialization.
In such a scenario, we might expect the line-of-sight component of the
velocities is biased for the same viewing angles as those for the projected
mass, projected pressure and projected galaxy number density.
This would lead to correlations in the mass inferred by richness, dynamics,
Compton distortion and lensing.
In this section we consider the relative strengths of these correlations and
their causes in the local cluster environment.

\subsection{Basic results}

\begin{figure}
\begin{center}
\resizebox{3in}{!}{\includegraphics{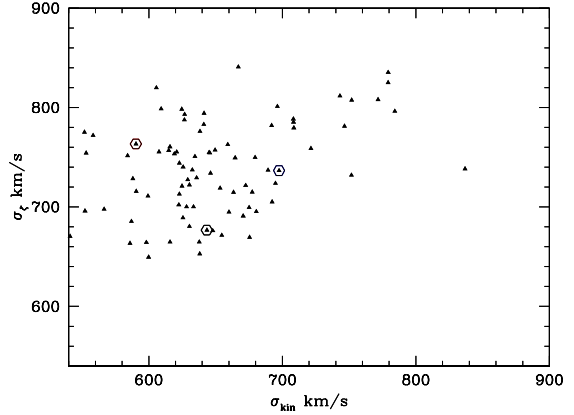}}
\end{center}
\caption{Correlated scatter for an individual cluster between velocity dispersion $\sigma_{\rm kin}$ and
lensing dispersion $\sigma_{\zeta}$.  The circles correspond to the
three lines of sight shown in Fig.~\ref{fig:hist3} for the same cluster.  Of 
the three, the lowest
lensing dispersion corresponds to the bottom panel there (with hardly any nearby
interlopers), the largest lensing dispersion (and lowest $\sigma_{\rm kin}$) 
corresponds to the center panel with many low mass neighbors, and
the largest velocity dispersion corresponds to the top panel, with a high mass
nearby halo.  The isotropic velocity dispersion for this cluster is 690 km/s.
}\label{fig:lenvel} 
\end{figure}

Considering richness, dynamics, lensing and Compton distortion, we found that
the degree of correlation between different measures of cluster size varied
dramatically from cluster to cluster, each sampling a different, local, cosmic
web.
Taking the median covariances from all the massive clusters the largest
covariances were between red galaxy richness and all other quantities,
followed by covariances of velocity dispersion with the other probes.
In terms of scatter of $M_{\rm pred}/M_{\rm true} -1$, the tightest
correlations with mass in our measures was for Compton distortion and phase
space richness, followed by weak lensing, and then red galaxy richness and
velocity dispersion\footnote{The scatter in Compton distortion and projected
mass could be increased by material outside our box, which we have not
modeled.}.  We show in Fig.~\ref{fig:lenvel} the measures of lensing dispersion
and velocity dispersion along all 96 lines of sight 
for the same cluster as in Fig.~\ref{fig:hist3}:  a correlation can be seen.

The correlations between individual measurements were usually below $0.5$,
indicating that each additional observation is adding significant new
information about the mass/size of the cluster, with the lower dispersion
measures giving the tightest constraints.
It should be borne in mind though that the distributions were far from Gaussian,
and the mass function steeply falling, so errors should be interpreted with
care.

To compare the ensemble of multiwavelength measurements for all the lines
of sights for all the clusters, we fit mean power-law relations between
the observables and mass to convert each multiwavelength measure to a common
system (the ``predicted'' mass).
Then we divided up the sightlines into ``good'' and ``bad'' based on whether
$|(M_{\rm true}-M_{\rm pred})|/M_{\rm true}\geq 0.5$ for at least 2 independent
observables\footnote{For this analysis we discarded lines of sight for clusters
where a higher mass cluster was found along the line of sight within $r_{180b}$
on the plane of the sky.  This takes out 70 out of our 7,968 massive sightlines.}.
The bad sightlines comprised 8 (11) per cent of the 96 sightlines per
cluster with $M\geq 2 (1)\times 10^{14}\,h^{-1}M_\odot$. 
Over half of the massive ($M\geq 2\times10^{14}\,h^{-1}M_\odot$) clusters had
at least one sightline where at least 3 measures were off
(the most common sources of mass errors were in red galaxy richness,
velocity dispersion and lensing), and more than half of the bad sightlines
were due to 18/83 of the clusters, each with 10 or more bad sightlines.
As $\sim 1/3$ of the sightlines had at least one quantity giving more than
a 50 per cent error from the mean mass relation, the reduction in error to
8 per cent of
the sightlines when using at least two measures is a significant improvement.
Using only galaxies with $L\geq 0.4\,L_\star$ (rather than $L\geq 0.2\,L_\star$)
resulted in a small increase in the number of bad sightlines
(from 8 to 11 per cent).

Scatter in the observables can arise from several violations of the
idealized, isolated, relaxed, spherical halo assumption.
The halo itself can be irregular (e.g.~recently merged), or regular but
anisotropic.  Nearby correlated material can be seen in projection or
uncorrelated material at large distances can be projected onto the cluster
position\footnote{We have tended to ignore this contribution here, as our
box is too small to fairly sample it and it has been extensively studied
elsewhere.}.
We have discussed halo state and anisotropy above.
Here our interest is in the comparison of nearby structure and substructure
for bad and good lines of sight.
We considered a cluster to have nearby massive structure if at least three
$L\geq 0.4\,L_\star$ galaxies from another halo(s) of mass
$\geq 0.2\,M_{\rm cluster}$ were present within $3\,\sigma_{\rm kin}$ in
redshift space and within $r_{180b}$ in the plane of the sky,
and to have nearby less massive structure if nearby massive structure
wasn't present as above and at least eight
$L\geq 0.4\,L_\star$ galaxies from halos with mass $\leq 0.2\,M_{\rm cluster}$
were within the same region.
For halos above $2 \times 10^{14}\,h^{-1}M_\odot$ we found 21 per cent of the
bad sightlines had nearby massive structure and 49 per cent had nearby less
massive structure, compared to 2 and 25 per cent respectively for the good
sightlines.  The bad sightlines were 10 times more likely to have a
nearby massive halo and almost twice as likely to have nearby less massive
halos.   
A larger fraction (52 per cent) of the bad sightlines have cluster
substructure (Dressler-Shechtman probability less than 0.05 as described in
\S\ref{sec:substructure}),  compared to 38 per cent of the good sightlines.
All together, 80 per cent of the bad sightlines had one of these three
indicators (nearby massive structure, or numerous less massive structure,
or substructure) compared to 51 per cent of the good sightlines.
These numbers changed very little when we lowered the mass threshold to
$10^{14} h^{-1} M_\odot$.  

However, although the likelihood of substructure, nearby massive or less massive halos increased for bad sightlines, the
majority of sightlines with substructure, nearby massive or less massive halos 
were not bad sightlines.
Of the 39 per cent of the lines of sight which
have substructure detected, only 10 per cent are bad lines of sight.
Similarly of the 4 per cent of sightlines with nearby massive structure,
59 per cent are good, and 41 per cent are bad.
For the 26 per cent with nearby less massive structure, 86 per cent are
good and 14 per cent are bad.

\subsection{Implications for stacking} \label{sec:stacking}

As is well known, correlated errors must be handled with care.
For example, if the source of scatter is correlated, two non-independent
measures can agree and both be in error.
These subtleties must also be borne in mind then one starts to stack
measurements \citep[see also][]{Nor08,Ryk08,Sta10}.

Stacking can be done in several ways.  Multiple measurements can be made for
a set of objects and then the mean of one of the measurements can be taken
holding another fixed.
Alternatively, there may be insufficient signal to measure all of the
properties on individual objects, so they are first stacked on one property
and the second is measured on the stack.
In this case one has the additional freedom to either scale the size
of any aperture with the first property or use a fixed metric aperture.
Finally, one can relate two properties while holding a third property fixed
either by averaging individual methods or measuring the properties on an
average (e.g.~fix richness and then measure Compton distortion and lensing).

It is known that a scatter between two variables, $x$ and $y$, implies that
conditional probabilities must be interpreted with care.
For example, there is scatter between halo mass, $M$, and richness, $N$,
which in the mean obey a relation lg$M=a+b\,{\rm lg}N$.
However the mean (log) mass of halos in a bin $N \approx N_0$ is not
$a+b\,{\rm lg}N_0$.
Since there are typically many more low mass halos than high mass halos,
it is likely that a high richness object is in reality a low mass object
with artificially high richness for its mass rather than an intrinsically
massive object of mean (or low) richness.  
The degree of such bias depends on the amount of scatter and the slope of the
halo mass function, which becomes steeper at both high mass and high redshift.
If one estimates the mass using a method (e.g.~lensing) which {\it itself\/}
has scatter, then the degree of error also depends on how correlated the
scatter between these methods is and the relative sizes of the scatter.

For example,  if scatter in richness were driven entirely by line-of-sight
projection of nearby structures, and if it was identical to the amount of
mass projected onto the ``lens'', then the error in the mass estimated by
lensing would cancel the bias described above.
However, if one measures an extra property, e.g.~X-ray flux, which is immune
to the projection, the mass--flux relation one infers from the stack would
be biased to high masses at fixed flux.
This would lead to an incorrect relation between mass and X-ray flux.
For detailed formulae in a simple analytical model see Appendix \ref{app:stack}
\citep[and][]{Ryk08,Nor08}.

In general we expect the situation to be slightly more complicated in reality
(or simulations) than the log-normal, analytical model suggests.  We saw in
the last section that a small number of halos are responsible for a fair
fraction of the outliers, and that the distribution of errors has non-Gaussian
tails.  While the general trends are not altered by these issues, they serve
to alter the quantitative predictions.

In fact none of these complications lead to large corrections to our measured
scaling relations.  All of the quantities show strong trends with mass, and
all of them have relatively large scatter.  The distribution of points in the
observable--observable plane is therefore determined by the range of masses
being selected much more than subtle correlations between the observables.
This serves to make any biases relatively small.
While we fully expect biases to be present, given our limited
simulation volume we are not able to measure them reliably.

Some examples serve to illustrate the main points.
We choose as a fiducial sample all lines-of-sight with red-sequence
richness $29\le N\le 30$ at $z\simeq 0.1$, containing 271 lines-of-sight from
104 halos.
We choose red-sequence richness as it is one of the more common quantities
to stack on.
As expected, the mean mass of these clusters is skewed low by the steeply
falling mass function.  The line-of-sight weighted mean mass is
$M_{180b}\simeq 2.2\times 10^{14}\,h^{-1}M_\odot$.  A randomly chosen sample
of halos with the same mass distribution has a line-of-sight weighted mean
richness of 22-24 (with fluctuations depending on how the sampling is done),
i.e.~it is $\sim 25$ per cent poorer than the input sample.
The mean (and median) values of the velocity dispersion, projected mass and
Compton distortion of this random sample are also ``low''.
How do these mean values compare to those of the sample selected on richness?
In fact they are quite similar, differing by $<10$ per cent in the mean.
This is because there is a large degree of scatter between red sequence
richness and mass and a strong correlation of all measures with mass,
making selecting on richness approximately the same as randomly sampling
halos with a specific mass distribution.
The joint distributions of e.g.~Compton distortion and velocity dispersion
or projected mass and Compton distortion also turn out to be very similar
in the random- and richness-selected samples.
There is a tendency for the richness-selected sample to have more
outliers in velocity dispersion using $3\sigma$ clipping than the
random sample, but otherwise the joint distributions are almost
indistinguishable.

We find similar results by stacking on e.g.~velocity dispersion.  The
distribution in e.g.~the $Y-\zeta$ plane is the same for the velocity
dispersion selected sample as in a sample of the same mass distribution.

The {\it largest\/} impact of stacking on e.g.~richness for our sample then is not
the degree to which the scatters in individual measurements are correlated
on an object-by-object basis but the fact that the stack contains clusters
of a wide range of masses/sizes.  If the measurement being performed is a
non-linear function of the mass, care must be taken in interpreting the
meaning of the averaged quantity.

\section{Higher redshift} \label{sec:highz}

Unfortunately our simulation volume is too small to make robust statements
about increasingly rare objects at high redshift, but in this section we
note some trends.  According to \citet{Mos10} the lower mass limit of our
subhalos corresponds to lower stellar-mass subhalos at higher $z$, with the
limit dropping from
$3\times 10^9\,h^{-1}M_\odot$ at $z\simeq 0$ to
$2\times 10^9\,h^{-1}M_\odot$ at $z\simeq 0.5$ to
$1\times 10^9\,h^{-1}M_\odot$ at $z\simeq 1$.
There is little evolution in the characteristic stellar mass in the
mass function over the same range, so we probe further below the break
in the mass function at higher $z$.
Since, on average, satellite subhalos fell into their host more recently
at higher $z$, the satellite fraction is smaller for samples selected
above a given halo mass or stellar mass
\citep[see discussion in e.g.][]{WetWhi10}.

While we have 83 halos with $M\geq2\times 10^{14}\,h^{-1}M_\odot$
at $z\simeq 0.1$, this drops to 28 at $z\simeq 0.5$ and only 5 at
$z\simeq 1$, making us increasingly susceptible to ``outliers''.
The number of massive neighbors per very massive halo increases as
we go to higher $z$, due to the steepening of the mass function at
the high-mass end.
Though the statistics are poor, there is evidence that the velocity bias
of the subhalos is decreasing with increasing redshift
\citep[see also][]{Evr08}.
The distribution of the eigenvalues of the velocity ellipsoid is very
similar to that at $z\simeq 0.1$ (shown in Fig.~\ref{fig:lamdist}),
again leading to large changes in line-of-sight velocity dispersion with
viewing angle.

At $z\simeq 0.5$ we found again that $M_{\rm kin}\propto R_h\sigma^2$ was
more tightly correlated with halo mass than $\sigma^3$, as it was at
$z\simeq 0.1$.
The more complex, phase-space interloper rejection method continued
to perform better than pure $3\,\sigma$ clipping.
In fact the trends of errors and correlations between mass and phase
space richness, Compton distortion, projected mass and velocity dispersion
were unchanged.  The phase-space richness and Compton distortion had the
least dispersion, followed by projected mass and then galaxy
kinematics\footnote{We did not consider red galaxy richness at higher
redshift as the method we used to assign color \citep{SkiShe09} was
only calibrated by observations for lower redshifts.}.
The fraction of bad sightlines does not change substantially going from
$z\simeq 0.1$ to $z\simeq 0.5$, however the fraction of these bad sightlines
with many interlopers from low-mass halos decreases.  As expected from the
increasing halos biases at higher redshift, the distribution of number of halos
around the massive clusters tended to have a higher mean at higher redshift.
The substructure fraction between $z=0.1$ and $z=0.5$ was close to unchanged.

\section{Conclusions}

Advances in observational capabilities and a new generation of wide-field
surveys have led to an explosion in multi-wavelength samples of galaxy
clusters.  By studying a cluster in many different wavebands, and from
many different approaches, we can obtain complementary information about
the physical state of the clusters and mitigate the systematic errors in
any single measurement.  Combining different measurements of cluster
properties has to be done carefully however, because the environment in
which clusters form leads to features which can be correlated across
methods.  As the correlation is not perfect, such a combination will
provide improvements over any individual method if done correctly.

In this paper we have used high resolution N-body simulations of a
cosmological volume to study how the large-scale environment of clusters
leads to correlated scatter in measures of cluster size, specifically those
based upon richness, Compton distortion, lensing or velocity dispersions.
Our simulation has enough force and mass resolution to track the subhalos
which are expected to host galaxies, allowing us to study dynamical probes
of the cluster with realistic samples incorporating a hierarchy of
substructures and retaining correlations between subhalo positions,
velocities and environment.
For this reason we paid particular attention to dynamical probes of
cluster size.

As might be expected in hierarchical structure formation scenarios, groups
of subhalos retain their identity for long periods within larger host halos.
This leads to a ``lack of virialization'' which implies that substructures
can thus behave quite coherently in phase space.
The highly anisotropic nature of infall into massive clusters, and their
triaxiality, means that line-of-sight galaxy velocity dispersions (or virial
masses) for any individual halo show large variance depending on viewing angle.
This suggests that dispersion-based mass estimators will work better in an
ensemble sense than for any individual cluster and that obtaining more than
tens of redshifts for any given cluster will not reduce the inferred mass
error.  We discussed the effect interloper galaxies, and their removal, has
on kinematic measurements and compared different schemes for interloper removal.
Results were presented both for individual clusters and for a ``stacked''
ensemble cluster.

All of our simulated clusters contain highly evident substructure, with
groups of subhalos which fall in together moving in a coherent fashion for
several Gyr.  However standard substructure indicators frequently miss this
substructure, and often give very different answers for a single object
viewed from different directions.
These results suggest caution when interpreting lack of observed
substructure in the galaxy distribution as evidence for a dynamically
relaxed, steady-state object (e.g.~justifying the use of the virial theorem
or Jeans analysis without the time derivative).
A cluster can be undergoing substantial mass accretion, i.e.~be far from
steady state, and still not be seen to have substructure along some lines
of sight.
The viewing angle dependence also complicates inferences about incidence of
dynamical evolution of cluster galaxies from observed interactions of
subclusters within the cluster identified through substructure finding
techniques.

Finally we note that many observational probes of clusters suffer from
projection effects, and that these are exacerbated by the dense, active
and anisotropic environments surrounding these massive objects.
We found increased nearby massive and less massive
halos, and substructure, when two of our measures (richness, lensing, 
Compton distortion and velocity dispersions) simultaneously had large outliers 
in predicted mass.  The converse was not always true, scatter in environment or
the measurement of substructure did not necessarily imply large outliers.

Since the orientation of the velocity ellipsoid is correlated with the
large-scale structure, velocity outliers also correlate with projection
induced outliers.  For many cases the same structure causes scatter in different
observations:
such scatters can be substantially correlated and this correlation needs to
be properly incorporated when combining measurements.

\bigskip

We would like to thank J. Bullock, A. Evrard, B. Gerke, H.Hoekstra, E. Rozo,
E. Rykoff, C. Stubbs, A. Wetzel and  A. Zabludoff for conversations and
A. Evrard and A. Wetzel for comments on the draft.
We thank A.~Zabludoff for suggesting we consider the ``virial mass''
$R_h\sigma^2$ in addition to $\sigma^3$.
M.W.~thanks Charlie Conroy and James Gunn for useful conversations and
collaborations about the stellar population synthesis technique.
We thank the referee, Andrea Biviano, for a constructive report and
helpful suggestions.
The simulations used in this paper were performed at the National Energy
Research Scientific Computing Center and the Laboratory Research Computing
project at Lawrence Berkeley National Laboratory.
M.W. was supported by the DoE and NASA.
R.S. was supported by the University of California Education Abroad 
Program.
J.D.C. thanks LBNL for travel support to the SNOWCLUSTER meeting and thanks its
organizers and participants for the opportunity
to present this work and for their suggestions and questions.

\appendix

\section{Finding subhalos}
\label{app:fof6d}

In our past work we have used the {\sl Subfind\/} algorithm \citep{SWTK} to
find subhalos.  However we have found that a phase-space based approach
performs better at tracking the subhalos in our most massive hosts and for
this reason we have switched to this new scheme here.
In particular we follow \citet{DieKuhMad06} and implement a phase-space
friends-of-friends finder.  Detailed experimentation, including a one-to-one
comparison of the new finder with the results of {\sl Subfind\/}, suggest that
choosing the configuration-space linking length to be $0.078$ of the mean
interparticle spacing and the velocity linking length to be $e^{-1}\simeq0.368$
of the halo (1D) velocity dispersion gives a good subhalo catalog.  As
discussed in \citet{DieKuhMad06} the results are stable to modest changes in
these parameters.  The most massive subhalos are the same for both finders,
but the lower mass structures which pass close to the center of the halo are
more robustly tracked in the phase-space method than with {\sl Subfind}.
We keep all 6D FoF halos which contain more than 20 particles.
For technical, book-keeping reasons if fewer than two subhalos (i.e.~a central
and a satellite) are found in any host halo we slowly increase the linking
lengths in that halo until one or two subhalos are found.
This ensures that there are no low-mass halos which have no subhalos,
simplifying the book-keeping in the tracking scheme.
This affects only the very low mass halos which are not used in this paper.

\section{The red sequence}
\label{app:color}

It has become common to use the tight red sequence of galaxies found in
clusters in order to isolate putative cluster members from chance alignments
along the line-of-sight during cluster detection.
The evolution of the red sequence with redshift means that choosing red
galaxies within a certain color cut also tends to give galaxies at a certain
redshift.  Because color-based cluster finders have become so popular, we
have included a toy-model of a color-based richness in our mocks.  There are
two steps, first to assign colors to the mock galaxies and second ask how the
observed properties depend on redshift/distance.
We take each of these in turn.

\subsection{Color assignment}

We first put colors into our $z\simeq 0.1$ box using the method of
\citet{SkiShe09}.  Their approach has red and blue galaxy populations
being drawn from two different populations (each with an
$M_r$-dependent mean and scatter), where the probability of a galaxy
belonging to either population depending upon $M_r$ and whether it
is a central or satellite galaxy.

We associate $r$-band magnitude with infall mass by abundance matching,
ignoring any scatter in the $M_r-M_{\rm inf}$ relation for simplicity.
\citet{SkiShe09} approximate the probability of a satellite to be red as
\begin{equation}
  P_{\rm sat}(M_r) =
  \frac{\langle g-r|M_r\rangle_{\rm sat} -\langle g-r|M_r\rangle_{\rm blue}}
  {\langle g-r|M_r\rangle_{\rm red} -\langle g-r|M_r\rangle_{\rm blue}}
\end{equation}
where
\begin{equation}
\begin{array}{lll}
\langle g-r|M_r \rangle_{\rm sat} & = & 0.83-0.08 (M_r + 20)  \\
\langle g-r|M_r \rangle_{\rm red} & = & 0.93-0.03 (M_r + 20) \\
\langle g-r|M_r \rangle_{\rm blue} & = & 0.62-0.11 (M_r + 20)
\end{array}
\end{equation}
and find an overall red fraction
\begin{equation}
  f_{\rm red}(M_r) \simeq 0.54 - 0.07 (M_r+20) \quad .
\end{equation}
Given $P_{\rm sat}(M_r)$, $f_{\rm red}$ and $f_{\rm sat}$
one can solve for $P_{\rm cen}(M_r)$, see Table \ref{tab:color}.
One then takes every galaxy, satellite or central, and randomly
decides whether it is red or blue.  If needed its actual color can
be taken from the Gaussian fits to the color-magnitude relations
given by \citet{SkiShe09}.

\begin{table}
\begin{center}
\begin{tabular}{ccccc}
 lg$M_{\rm inf}$ & $M_r$ & $f_{\rm red}$ & $P_{\rm cen}$ & $P_{\rm sat}$ \\
 \hline
11.50 & -19.1 &  0.48 &  0.41 &  0.62 \\
12.00 & -20.2 &  0.56 &  0.50 &  0.69 \\
12.50 & -20.9 &  0.60 &  0.56 &  0.76 \\
13.00 & -21.4 &  0.64 &  0.60 &  0.83 \\
13.50 & -21.8 &  0.67 &  0.63 &  0.91
\end{tabular}
\end{center}
\caption{Magnitudes and red fractions as a function of infall mass (in
$h^{-1}M_\odot$) from \protect\citet{SkiShe09}.  $f_{\rm red}$ is the
fraction of all galaxies which are red, while $P_{\rm cen}$ and
$P_{\rm sat}$ are the probabilities that a central or satellite galaxy
of that infall mass is red.}
\label{tab:color}
\end{table}

\subsection{Evolution with redshift}

The fact that the observed colors of galaxies evolve with distance means
that a tight sequence in color (e.g.~the red sequence) will shift out of
any thin color slice as the galaxies shift in distance.  Thus cuts in color
can be used to isolate galaxies within a small range of distances
\citep[e.g.][]{RCSI,RCSII}.
Modeling the evolution of galaxy colors {\it ab initio\/} is notoriously
difficult, but a hybrid method based on stellar population synthesis models
can isolate the main features for our purposes of making ``pseudo'' light cones.

We simplify our problem by assuming that blue galaxies can be distinguished
{}from red at any distance and we need only consider the evolution of the
red galaxies.  We make the further simplification that all of the red
galaxies are evolving passively, with star formation having ceased at some
high redshift (e.g.~$z\ge 2$).
Using the stellar population models described in \citep{SP1,SP2,SP3}
we find that the $g-r$ color of a passively evolving population scales with
redshift as $d(g-r)/dz\simeq 1$--2, with the precise slope depending on
the star formation history assumed.  Similarly, the absolute $r$-band magnitude
scales as $dM_r/dz\simeq 0.1$--1.  For our cosmology $d\chi/dz=2900\,h^{-1}$Mpc,
where $\chi$ is the (comoving) line-of-sight distance and a linear approximation
is acceptable over the limited extent of our simulation.

Given a color cut of a certain width the speed at which the color of the
red sequence ``ridgeline'' changes with $z$ defines the range of distance over
which galaxies will be selected\footnote{Even though our box is at a single
output time, we assume line-of-sight distance corresponds to redshift.  The
evolution of the large-scale structure over the relevant time interval is so
small that it may be safely neglected.}.
We encode this information as the probability that a red galaxy at a given
distance will fall into the fiducial red sequence cut
\citep[c.f.][]{Coh07}.

If the width and peak of the red sequence were independent of the magnitude
this transformation would be trivial: for a Gaussian color distribution and a
fixed width $\Delta c$ the interloper probability is the difference of two
error functions with width $\Delta c/(dc/d\chi)$.
However the non-zero dependence on $M_r$ slightly complicates the calculation.
To include this complication we first calculate the corresponding magnitude
for every red galaxy as if it were actually at redshift $0.1+\delta z$,
corresponding to its offset from the box midplane, but dimmed (or brightened)
by the change in distance.  
We then calculate what $M_r$ and thus $g-r$ distribution will result for
this dimmed galaxy as it evolved to $z=0.1$, assuming the linear evolution
defined above.
The evolved $M_r$ at $z=0.1$ has a $z=0.1$ color ($g-r$) distribution well
fit by a Gaussian distribution \citep{SkiShe09}.
The color is evolved back to $z=0.1+\delta z$ to give the observed color
distribution for the galaxy at $z=0.1+\delta z$ with magnitude $M_r$ estimated
as if it were at $z=0.1$.
The interloper fraction of galaxies is the integral of the observed
distribution within the cut defining the red sequence.

If we make our red sequence selection have $g-r$ width 0.05 we find the
dispersion in distance runs between $50$ and $100\,h^{-1}$Mpc, depending
on stellar population model, galaxy magnitude etc.  To be conservative, in
the sense of reducing line-of-sight projection, we take the lower end of
the range and assume a red galaxy in the foreground or background of the
cluster of interest is included within the red sequence with a Gaussian
probability of width $50\,h^{-1}$Mpc.  As with the other measures, we
apodize the selection to ensure the probability is zero at the limits of
the simulation.

\section{Measurements with correlated scatter and stacking}
\label{app:stack}

It is helpful to consider a simple analytic model which illustrates
the effect of correlated scatter on different observables.  We will
consider the case of two measurements, $m_1$ and $m_2$, of some quantity
$m$.  For example, one could consider $m_1$ to be richness-inferred (log) mass
and $m_2$ to be lensing-inferred (log) mass with $m$ the ``true'' (log) mass.
We imagine that $P(m_1,m_2|m)$ is a bi-variate Gaussian with means
$\mu_i(m)$ and covariance
\begin{equation}
  {\rm Cov}\left[m_1,m_2\right] = \left\{
    \begin{array}{cc}
       \sigma_1^2 &   \rho_{12} \sigma_1 \sigma_2 \\
      \rho_{12} \sigma_1 \sigma_2  &   \sigma_2^2 \\
    \end{array}
  \right\} \quad .
\end{equation}
and for simplicity we assume that $\sigma_i$ and $\rho_{12}$ are independent
of $m$ and that $\mu_i=a_i+b_im$.
We shall write the probability that a cluster has mass $m$, the mass function,
as $P(m)$ and for convenience/simplicity take it to be a power-law in mass or
$P(m)\propto \exp[-\alpha m]$.

As is well known, when $\sigma_i>0$ and $\alpha\ne 0$ the mean ``true''
mass of clusters with measured mass $m_i$ is biased.  Since
$P(m|m_1) \propto P(m_1|m)P(m)$ we have
\begin{equation}
  \left\langle m|m_1=m_1^{\rm obs}\right\rangle
  \equiv \bar{m}
  = \frac{m_1^{\rm obs}-a_1}{b_1} - \frac{\alpha}{b_1^2}\,\sigma_1^2
  \quad .
\end{equation}
Similarly, if we consider the case where $m_1$ is known (e.g. selected) to be $m_1^{\rm obs}$
the conditional distribution of $m_2$ is also a Gaussian with
$\sigma_2'=(1-\rho_{12}^2)\sigma_2^2$ and mean
\begin{equation}
  \left\langle m_2|m_1=m_1^{\rm obs}, m\right\rangle =
    \mu_2 + \frac{\sigma_2}{\sigma_1} \rho_{12}
    \ \left(m_1^{\rm obs}-\mu_1\right)
  \quad .
\label{eqn:condmean}
\end{equation}

These facts allow us to consider computing $m_2$ as a function of $m$ by
binning on $m_1$ and averaging the measures of $m_2$ in each bin.
It is easy to show that if $\sigma_1=0$ one simply obtains
$\langle m_2\rangle=a_2+b_2\bar{m}=a_2+(b_2/b_1)(m_1^{\rm obs}-a_1)$
as desired.
However if the $\sigma_i>0$ we have more terms.  By writing
$P(m_1,m_2,m)=P(m_2|m,m_1)P(m|m_1)P(m_1)$ and recalling that the
$\mu_i$ are linear in $m$ one finds Eq.~(\ref{eqn:condmean}) with
$\mu_i(\bar{m})$ in place of $\mu_i(m)$ leading to
\begin{equation}
  \left\langle m_2\right\rangle = a_2
  + \frac{b_2}{b_1}\left(m_1^{\rm obs}-a_1\right)
  + \frac{\alpha}{b_1}
  \left(\rho_{12}\sigma_1\sigma_2-\frac{b_2}{b_1}\sigma_1^2\right)
  \, .
\label{eqn:m2m1}
\end{equation}

Consider the case $a_i=0$ and $b_i=1$, i.e.~the measurements give unbiased
estimates of the (log) mass for halos of a fixed mass:
$\left\langle m_i|m\right\rangle=m$.
Stated another way, the average of each mass estimate in narrow bins of halo
mass returns the correct mass and in such bins each measurement correctly
predicts the mass which would be estimated by every other measurement.
If one could bin in mass, it would be straightforward to estimate the mean
observable-mass relation.

The situation changes when we bin not by mass but by observable, e.g.~richness.
In this case, even though the richness-based mass estimator is unbiased, the (unobservable)
true mean halo mass in the bin is biased (low) because the falling mass
function makes it more likely that a halo of richness $N$ is a low mass halo
which fluctuated up than a high mass halo which fluctuated down in richness.
Similarly, the mean mass estimated from a second observable in that bin
differs by $\alpha(\rho_{12}\sigma_1\sigma_2-\sigma_1^2)$ from the first
observable defining the bin, as in Eqn. ~\ref{eqn:m2m1}.
That observable-mass relation is thus also biased.
Note that the bias disappears if $\rho=1$ and $\sigma_1=\sigma_2$, in which
case the errors conspire to cancel exactly because fluctuations in observable
one directly imply compensating fluctuations in observable two.
For example, a low mass halo which had an abnormally high richness would
be counted in the richness bin even when it ``should not be''.  But its
lensing signal would also be abnormally high by just the right amount to
give the right mean mass in the richness bin.

Finally, if we estimate a third observable in the same bins of e.g.~richness
it will be biased by a different amount:
$\alpha(\rho_{13}\sigma_1\sigma_3-\sigma_1^2)$.
The relation between observables $2$ and $3$, when binned on $1$, is thus
biased in both coordinates.
Though we have not considered it in our toy model, this bias may well be
mass dependent.

In these examples the bias is due entirely to the falling mass function,
because we assumed explicitly that $a_i=0$ and $b_i=1$, i.e.~the measurements
give unbiased estimates of the mass.  Using the results above, the general
case can be considered but we gain no further insight from doing so.

\label{lastpage}

\end{document}